\documentclass{article}

\usepackage{arxiv}

\usepackage[utf8]{inputenc} 
\usepackage[T1]{fontenc}    
\usepackage[colorlinks=true, linkcolor=red, citecolor=green, urlcolor=blue]{hyperref}
\usepackage{url}            
\usepackage{booktabs}       
\usepackage{amsfonts}       
\usepackage{nicefrac}       
\usepackage{microtype}      
\usepackage{lipsum}		
\usepackage{graphicx}
\usepackage{natbib}
\usepackage{optidef}
\usepackage{doi}
\usepackage{algorithm}
\usepackage{algpseudocode}
\usepackage{xcolor}

\usepackage{amsmath,amssymb}

\usepackage[T1]{fontenc}
\usepackage{bm}  

\usepackage{xcolor}
\usepackage{hyperref}

\hypersetup{
    colorlinks=true,
    linkcolor=blue,    
    citecolor=blue,   
    urlcolor=cyan      
}

\usepackage{graphicx}

\newcommand{\g}{{\bm{g}}}
\newcommand{\D}{\bm{D}}
\newcommand{\m}{\bm{m}}
\renewcommand{\v}{\bm{v}}

\renewcommand{\d}{{\bm{d}}}
\newcommand{\z}{{\bm{z}}}
\newcommand{\G}{{\bm{G}}}
\newcommand{\W}{\bm{W}}
\renewcommand{\r}{\bm{r}}
\renewcommand{\m}{\bm{m}}
\newcommand{\n}{\bm{n}}
\newcommand{\J}{\bm{J}}
\newcommand{\A}{\bm{A}}
\newcommand{\B}{\bm{B}}
\newcommand{\C}{\bm{C}}
\newcommand{\I}{\bm{I}}
\newcommand{\x}{\bm{x}}
\newcommand{\R}{\bm{R}}
\newcommand{\Q}{\bm{Q}}

\title{Adaptive Seismic Inversion in Piecewise Smooth Media Using Whiteness-based Alternating Directions}

\date{} 					

\author{
	Lian Zhao\\
	the State Key Laboratory of Deep Oil and Gas\\
	China University of Petroleum (East China)\\
	Qingdao, China \\
	\texttt{zhaol\_yt@upc.edu.cn} \\
	\And
	Kamal Aghazade \\
	Institute of Geophysics\\
	Polish Academy of Sciences\\
	Warsaw, Poland \\
	\texttt{aghazade.kamal@igf.edu.pl} \\
    \And
	Ali Gholami \\
	Institute of Geophysics\\
	Polish Academy of Sciences\\
	Warsaw, Poland \\
	\texttt{agholami@igf.edu.pl} \\
	\And
	Danping Cao \\
	the State Key Laboratory of Deep Oil and Gas\\
	China University of Petroleum (East China)\\
	Qingdao, China \\
	\texttt{caodp@upc.edu.cn} \\
}

\renewcommand{\shorttitle}{} 

\hypersetup{
pdftitle={Adaptive Seismic Inversion in Piecewise Smooth Media Using Whiteness-based Alternating Directions},
pdfsubject={q-bio.NC, q-bio.QM},
pdfauthor={David S.~Hippocampus, Elias D.~Striatum},
pdfkeywords={First keyword, Second keyword, More},
}

\begin{document}
\maketitle

\begin{abstract}
Post-stack acoustic impedance inversion is ill-posed because seismic data are band-limited and contaminated by noise. Regularization improves the stability of the inversion, but Tikhonov regularization tends to blur formation interfaces, whereas total variation (TV) regularization may introduce staircase artifacts. To simultaneously characterize abrupt interlayer discontinuities and gradual intralayer variations, we propose an inversion method based on anisotropic Tikhonov–TV (ATT) regularization and automatic parameter selection. The method decomposes the logarithmic impedance model into piecewise-constant and smooth components. TV and Tikhonov constraints are imposed on these components, respectively, in both the vertical and horizontal directions. The performance of the inversion depends critically, among other factors, on the relative weight of the total regularization term with respect to the data misfit (the regularization parameter), the relative weight of the two regularization components (the balancing parameter), and the efficiency of computing a candidate solution for a given set of parameters. In this paper, we propose a new adaptive algorithm based on alternating directions, in which (i) a fast candidate-solution computation scheme is developed in the spectral domain by exploiting the Sylvester structure and fast Fourier transforms, thereby reducing the computational cost of parameter searching; (ii) the regularization parameter is selected adaptively using the residual whiteness principle (RWP); and (iii) the balancing parameter is adaptively updated according to robust statistical characteristics of the impedance-model gradients. Synthetic and field-data tests demonstrate that, compared with conventional Tikhonov and TV methods, the proposed method better preserves impedance interfaces while reducing staircase artifacts and inter-trace discontinuities, thereby improving the reconstruction of impedance structures.
\end{abstract}

\section{Introduction}
Post-stack seismic impedance inversion estimates the subsurface acoustic impedance distribution from band-limited post-stack seismic records and supports reservoir characterization and quantitative seismic interpretation. Limited bandwidth and noise make the inversion ill-posed \cite{Lindseth_1979_SSL,Oldenburg_1982_RAI,Cooke_1983_GLI,Zong_2012_AVO}. Therefore, low-frequency models and prior constraints are commonly incorporated to render the solution unique and stable \cite{russell_1988_introduction,ghosh_2000_limitations,wu2017structure}.

Tikhonov regularization is a classical approach to stabilizing ill-posed inverse problems \cite{tikhonov1977solutions}. By penalizing the $\ell_2$-norm of the model or its derivatives, this method suppresses noise amplification and unstable oscillations, while offering the advantages of a simple formulation, high computational efficiency, and good numerical stability. In impedance inversion, Tikhonov regularization can effectively recover the smooth background variations in impedance. However, the quadratic penalty strongly penalizes large gradients, including those associated with genuine impedance interfaces, and may therefore blur layer boundaries \cite{chopra2009relative,zhang2022novel}.

To enhance the interface-preserving capability of inversion results, sparse regularization has increasingly been introduced into inversion. These methods exploit the sparsity of reflection series or gradients and improve the vertical resolution of impedance inversion through the $\ell_1$-norm and its generalized forms \cite{sacchi1997reweighting,wang2010seismic}. Among them, total variation (TV) regularization constrains the $\ell_1$-norm of the model gradient, thereby preserving impedance discontinuities while suppressing noise \cite{rudin1992nonlinear}. Gholami \cite{gholami2015nonlinear} applied TV regularization to multichannel impedance inversion and improved the quality of the impedance estimates. Wang et al. \cite{wang2018data} subsequently introduced anisotropic TV constraints to enhance the spatial continuity of the inversion results. Higher-order TV and nonconvex TV methods were later developed to improve structural characterization and alleviate the limitations of conventional TV regularization \cite{bredies2010total,wu2019seismic,zou2026post,zhang2026total}. However, first-order TV regularization promotes piecewise-constant reconstructions and can therefore introduce staircase artifacts in smoothly varying impedance regions. It may therefore excessively suppress continuous variations within formations, causing smoothly varying impedance to be reconstructed as multiple constant intervals and consequently producing staircase artifacts. 

Subsurface impedance commonly contains both abrupt interlayer discontinuities and gradual intralayer variations. Tikhonov regularization is well suited to characterizing continuous variations, whereas TV regularization is effective in preserving sharp interfaces. However, either constraint alone cannot adequately account for both structural characteristics. To overcome the limitations of single regularization methods, researchers have explored compound regularization strategies. Gholami and Hosseini \cite{gholami2013balanced} decomposed the model to be estimated into a smooth component and a piecewise-constant component, which were constrained by second-order Tikhonov regularization and first-order TV regularization, respectively, thereby establishing the compound Tikhonov–TV (TT) regularization framework. This method avoids restricting the entire model to either a purely smooth or a blocky structure. However, the resulting model decomposition is sensitive to the balancing parameter between the two regularization terms. To reduce dependence on a predefined parameter, Gholami and Gazzola \cite{gholami2022automatic} further developed an adaptive balancing strategy based on the median absolute deviation (MAD) and robust z-scores of the model gradients. Subsequently, this adaptive strategy was applied to acoustic and elastic full-waveform inversion, demonstrating favorable adaptability in the joint recovery of smooth backgrounds and sharp interfaces \cite{aghazade2024elastic,aghazade2026robust}.

However, existing TT methods mainly focus on balancing the smooth and piecewise-constant components, and have not fully considered the variation differences of the impedance structure in different directions. In practice, impedance not only exhibits piecewise-smooth behavior in the vertical direction, characterized by gradual intralayer variations and abrupt interlayer discontinuities, but also displays overall continuity accompanied by localized variations in the horizontal direction. If only the vertical variation is considered, it is difficult to effectively suppress inter-trace discontinuity artifacts, thereby compromising the accurate recovery of impedance structures. It is therefore necessary to jointly constrain smooth variations and impedance discontinuities in both directions.

In addition to the form of regularization, associated parameter selection affects both inversion stability and computational efficiency. This parameter must be selected properly to balance between the total regularization and the data misfit. Conventional trial-and-error tuning is experience dependent and time consuming, making effective regularization parameter estimation strategies essential for solving the inversion problem. Previous studies on TT regularization have reduced the dependence of the regularization parameter on a predefined value, often requiring the noise level. However, when the optimization problem is solved using the alternating direction method of multipliers (ADMM) \cite{esser2009applications,boyd2011distributed,ghadimi2014optimal}, this parameter must still be selected appropriately for an improved convergence. For regularization parameter selection, Almeida and Figueiredo \cite{almeida2013parameter} proposed evaluating reconstruction results using the residual whiteness principle (RWP). Pragliola et al. \cite{pragliola2023admm} applied the RWP to parameter selection for image super-resolution within the ADMM framework, and Aghazadeh et al. \cite{aghazade2025automatic} further extended it to penalty parameter selection in full-waveform inversion. These studies provide a basis for selecting the regularization parameter by minimizing the kurtosis of the Fourier-transformed residuals. Within the ADMM framework, the kurtosis function must be evaluated over a range of candidate parameter values at each iteration to identify its minimizer \cite{pragliola2023admm,aghazade2025automatic}. However, in the impedance inversion formulation considered here, separately solving the model subproblem for multiple candidate parameter values at each iteration would incur a high computational cost. Fast computation of the candidate solutions is therefore required to efficiently incorporate RWP-based parameter selection into the inversion iterations.

In this paper, we propose a post-stack acoustic impedance inversion method using anisotropic Tikhonov–TV (ATT) regularization and automatic parameter selection. The proposed method separately constructs vertical and horizontal difference terms and jointly imposes Tikhonov and TV constraints along both directions to characterize smooth intralayer variations and localized impedance discontinuities. The two parameters, controlling the balance between data fitting and regularization and the relative balance between the two regularization terms, respectively, are jointly updated at each iteration of the alternating-direction algorithm. For the first time, this study combines two statistical measures, the kurtosis score and the robust z-score, to adaptively determine these two parameters at each iteration. To make the proposed algorithm computationally practical, we develop an efficient spectral-domain solver for the main subproblem by exploiting its Sylvester-equation structure and fast Fourier transforms.
Synthetic model and field-data tests demonstrate that, compared with Tikhonov, TV, and the standard TT regularization methods, the proposed ATT method achieves a better balance between interface delineation and continuous intralayer variations and improves the convergence and accuracy of the inversion results.

\section{Method}
\subsection{Forward modeling}
Using the convolutional model \cite{robinson1967predictive}, the post-stack seismic data $\d \in \mathbb{R}^{N_t \times N_x}$ are written as: 
\begin{equation}
 \d = \W \r + \n, 
\label{equation:1}
\end{equation}
where $\W \in \mathbb{R}^{N_t \times N_t}$ denotes the wavelet convolution matrix, $\r \in \mathbb{R}^{N_t \times N_x} $ denotes the reflectivity series, and its last row is set to zero, and $\n \in \mathbb{R}^{N_t \times N_x} $  represents noise. 

For small logarithmic impedance differences between adjacent samples, the reflection coefficient can be approximated as:
\begin{equation}
    \r_t = \frac{\m_{t+1}-\m_{t}}{2},
    \label{equation:2}
\end{equation}
where $\m=\text{ln}(\z) \in \mathbb{R}^{N_t \times N_x}$ denotes logarithm of the impedance, and $\z$ represents impedance.

Substituting equation (\ref{equation:2}) into equation (\ref{equation:1}) yields:
\begin{equation}
\d=\G \m+\n,
\label{equation:3}
\end{equation}

where $\G=\frac{1}{2} \W \D$ denotes the forward operator, and $\D$ denotes the first-order difference operator.
\begin{equation}
\D=\left[\begin{array}{ccccc}
-1 & 1 & & &  \\
& -1 & 1 & & \\
& & \ddots & \ddots & \\
& & & -1 & 1 \\
& & & & 0
\end{array}\right]_{N_t \times N_t}
\label{equation:4}
\end{equation}

Equation (\ref{equation:3}) shows that seismic data primarily constrain the wavelet-filtered impedance differences. Because the difference operator cannot determine the constant component of the model and the band-limited nature of the wavelet weakens the constraints on certain frequency components, it is difficult to stably recover the absolute impedance and its background trend through data fitting alone. Therefore, both background information and regularization constraints consistent with the structural characteristics of impedance must be incorporated:
\begin{equation}
\underset{\m}{\operatorname{minimize}}~ \frac{1}{2\mu}\|\d-\G \m\|_F^2+\frac{1}{2\rho}\left\|\m-\m_{\mathbf{0}}\right\|_F^2+R_\beta(\m),\ 
\label{equation:5}
\end{equation}
where $\|\cdot\|_F^2 $ denotes the Frobenius norm, $\rho$ denotes the weight of the low-frequency background constraint. $\m_0$ represents the known low-frequency model. $R_\beta(\m)$ represents the regularization constraint by $\beta >0$, and $\mu >0$ is the total regularization parameter that balances between the fit to the data and the regularization term. While the inversion algorithm is robust to the value of $\rho$, it is sensitive to both $\mu$ and $\beta$. Thus, their value must be determined properly for the optimum performance of the inversion.

\subsection{The anisotropic Tikhonov-TV regularization}

The structural regularization term $R_\beta (\m)$ encodes prior assumptions about the impedance model \cite{asnaashari2013regularized,guitton2012blocky,guitton2012constrained}. In practice, impedance generally exhibits piecewise-smooth behavior, characterized by distinct impedance discontinuities at formation interfaces and predominantly smooth, continuous variations within formations. 

To represent these characteristics, the logarithmic impedance model is decomposed as:
\begin{equation}
 \m =\m_b + \m_s,
 \label{equation:6}
\end{equation}
where $\m_b$ denotes the piecewise-constant component that characterizes interlayer discontinuities, and $\m_s$ denotes the smooth component that describes continuous intralayer variations.

In a trace-by-trace inversion setting, the standard TT regularization applies a first-order TV penalty to $\m_b$ and a second-order Tikhonov penalty to $\m_s$, both applied along the vertical direction:
\begin{equation}
 R_\beta (\m = \m_b + \m_s) = \| \D_1 \m_b \|_1 + \frac{\beta}{2} \| \D_2 \m_s \|_F^2,
 \label{equation:7}
\end{equation}
where $\|\cdot\|_1$ denotes the entrywise $\ell_1$ norm (the sum of the absolute values of all entries in an argument matrix), $\beta$ controls the relative weighting between the smooth and piecewise-constant components. $\D_1=\D$ and $\D_2=\D_1\D_1$ denote the first- and second-order vertical difference operators, respectively.

Equation \eqref{equation:7} constrains impedance variations only in the vertical direction and does not exploit the lateral correlations between adjacent seismic traces. Considering that formations also exhibit a combination of overall continuity and localized variations in the horizontal direction \cite{zhao2025alternate}, we separately construct vertical and horizontal difference terms and impose first-order TV and second-order Tikhonov constraints along both directions, thereby extending conventional TT regularization to an anisotropic regularization form:
\begin{equation}
\begin{aligned}
R_\beta(\m)= & \|\D_{1} \m_{\boldsymbol{b}}\|_1+\left\|\m_{\boldsymbol{b}} \D_{1 x}\right\|_1   
 +\frac{\beta}{2}\left( \|\D_{2} \m_s\right\|_F^2+ 2\|\D_{1} \m_s \D_{1 x}\|_F^2+\left\|\m_s \D_{2 x}\|_F^2 \right),
\end{aligned}
 \label{equation:8}
\end{equation}
where $\D_{1 \boldsymbol{x}}$ and $\D_{2 \boldsymbol{x}}=\D_{1 \boldsymbol{x}} \D_{1 \boldsymbol{x}}$ denote the first- and second-order horizontal difference operators, respectively.
\begin{equation}
\D_{\mathbf{1} \boldsymbol{x}}=\xi\left[\begin{array}{ccccc}
-1 & & & & \\
1 & -1 & & & \\
& 1 & \ddots & & \\
& & \ddots & -1 & \\
& & & 1 & 0
\end{array}\right]_{N_x \times N_x}   
\label{equation:9}
\end{equation}
where $\xi$ is the anisotropy parameter.

Substituting \eqref{equation:8} into \eqref{equation:5} gives the anisotropic form of the TT-regularized objective function:

\begin{equation}
 \begin{aligned}
\underset{\m, \m_s, \m_b}{\operatorname{minimize}}~ \frac{1}{2\mu} \| \d-  \G \m \|_F^2 & + \frac{1}{2\rho} \| \m-\m_0 \|_F^2 
  + \| \D_1 \m_b \|_1 +\left\|\m_b \D_{1 x}\right\|_1 \\
& +\frac{\beta}{2}\left(\|\D_2 \m_s\right\|_F^2 + 2\left\|\D_1 \m_s \D_{1 x}\|_F^2+\|\m_s \D_{2 x}\|_F^2\right),\\
&\hspace{-3.8cm} \text{subject to }~~\m ~= ~\m_b + \m_s.
\end{aligned}
\label{equation:10}
\end{equation}
We introduce the auxiliary gradient variables $\g_{11} = \D_1 \m_b$, $\g_{12} = \m_b\D_{1x}$, $\g_{21} = \D_{1}\m_s$ , $\g_{22} = \m_s\D_{1x}$. Specifically, the variables $\g_{11}$ and $\g_{12}$ denote the first-order vertical and horizontal gradients of the piecewise-constant component $\m_b$, respectively, whereas $\g_{21}$ and $\g_{22}$ denote the first-order vertical and horizontal gradients of the smooth component $\m_s$, respectively. Accordingly, equation \eqref{equation:10} can be reformulated in terms of the gradient variables as:
\begin{equation}
 \begin{aligned}
\underset{\m, \g_{11},\g_{12},\g_{21},\g_{22}}{\operatorname{minimize}}~  \frac{1}{2\mu} \| \d-  \G \m \|_F^2&+  \frac{1}{2\rho} \| \m-\m_0 \|_F^2 
 +\| \g_{11} \|_1 +\left\|\g_{12}\right\|_1 \\
&+\frac{\beta}{2}\left(\|\D_1 \g_{21}\right\|_F^2
 +\left\|\g_{21}\D_{1 x}\|_F^2+ \|\D_1\g_{22}\|_F^2+\|\g_{22}\D_{1x}\|_F^2\right),\\
&\hspace{-4.3cm} \vcenter{\hbox{subject to}} \quad \begin{array}{rcl}
\D_1\m &=& \g_{11} + \g_{21}, \\
\m\D_{1x} &=& \g_{12} + \g_{22}.
\end{array}
\end{aligned}
\label{equation:11}
\end{equation}
\subsection{Solving the ATT-Regularized problem \eqref{equation:11} using ADMM}
To solve the optimization problem in equation \eqref{equation:11}, Lagrange multipliers are introduced, and the problem is expressed in augmented Lagrangian form:
\begin{equation}
\begin{aligned}
\J(\m, \g_{11},\g_{12},\g_{21},\g_{22},\v_1,\v_2) 
 &= \frac{1}{2\mu} \| \d- \G \m \|_F^2+ \frac{1}{2\rho} \| \m-\m_0 \|_F^2 
 +\| \g_{11} \|_1 + \|\g_{12}\|_1 \\
& +\frac{\beta}{2} \Bigl(\|\D_1 \g_{21}\|_F^2+\|\g_{21}\D_{1 x}\|_F^2 
 + \|\D_1\g_{22}\|_F^2+\|\g_{22}\D_{1x}\|_F^2 \Bigr) \\
& + \frac{1}{2\gamma} \Bigl( \| \D_1\m-\g_{11}-\g_{21}-\v_1 \|_F^2
 +  \| \m\D_{1x}-\g_{12}-\g_{22}-\v_2\|_F^2 \Bigr),
\end{aligned}
\label{equation:12}
\end{equation}
where $\v_1$ and $\v_2$ denote the Lagrange multipliers, and $\gamma$ denotes the penalty parameter.
In the scaled form ADMM iterations, the variables are updated sequentially through the following subproblems:
\begin{align}
\m^k &= \underset{\m}{\operatorname{argmin}}~ \frac{1}{2\mu} \| \d- \G \m \|_F^2+  
\frac{1}{2\rho} \| \m-\m_0 \|_F^2 \nonumber \\
&\quad + \frac{1}{2\gamma} \Bigl( \| \D_1\m-\g_{11}^{k-1}-\g_{21}^{k-1}-\v_1^{k-1} \|_F^2
 +  \| \m\D_{1x}-\g_{21}^{k-1}-\g_{22}^{k-1}-\v_2^{k-1}\|_F^2 \Bigr),
\label{equation:13} \\[1ex]
\g_{11}^k &= \underset{\g_{11}}{\operatorname{argmin}}~  \| \g_{11} \|_1 +  \frac{1}{2\gamma} \| \D_1\m^{k}-\g_{11}-\g_{21}^{k-1}-\v_1^{k-1} \|_F^2,
\label{equation:14} \\[1ex]
\g_{12}^k &= \underset{\g_{12}}{\operatorname{argmin}}~  \| \g_{12} \|_1 +  \frac{1}{2\gamma} \| \m^{k}\D_{1x}-\g_{12}-\g_{22}^{k-1}-\v_2^{k-1} \|_F^2,
\label{equation:15} \\[1ex]
\g_{21}^k &= \underset{\g_{21}}{\operatorname{argmin}}~  \frac{\beta}{2}(\|\D_1\g_{21}\|_F^2+\|\g_{21}\D_{1 x}\|_F^2) + \frac{1}{2\gamma}\| \D_1\m^{k}-\g_{11}^{k}-\g_{21}-\v_1^{k-1} \|_F^2,
\label{equation:16} \\[1ex]
\g_{22}^k &= \underset{\g_{22}}{\operatorname{argmin}}~  \frac{\beta}{2}(\|\D_1\g_{22}\|_F^2+\|\g_{22}\D_{1 x}\|_F^2) + \frac{1}{2\gamma}\| \m^{k}\D_{1x}-\g_{12}^{k}-\g_{22}-\v_2^{k-1} \|_F^2,
\label{equation:17} \\[1ex]
\v_1^k &= \underset{\v_1}{\operatorname{argmin}}~  \v_1^{k-1} + \g_{11}^{k} + \g_{21}^{k} - \D_1\m^{k},
\label{equation:18} \\[1ex]
\v_2^k &= \underset{\v_2}{\operatorname{argmin}}~  \v_2^{k-1} + \g_{12}^{k} + \g_{22}^{k} - \m^{k}\D_{1x}.
\label{equation:19}
\end{align}

In what follows we describe how to solve each subproblem efficiently.

The subproblem in \eqref{equation:13} is quadratic and convex in $\m$. Its first-order optimality condition has the form of a Sylvester equation \cite{gholami20193d,lin2023time,zhao20233}:
\begin{equation}
 \begin{aligned}
\A\m^k + \m^k\B= \C.
\end{aligned}
\label{equation:20}
\end{equation}
Given by
\begin{equation} \m^k=\operatorname{Sylvester}(\boldsymbol{A}, \boldsymbol{B}, \boldsymbol{C}),
    \label{equation:21}
\end{equation}
with
\begin{equation*}
\left\{\begin{aligned}
\boldsymbol{A}= & \boldsymbol{G}^T \boldsymbol{G}+\lambda \boldsymbol{D}_1^T \boldsymbol{D}_1+\alpha \boldsymbol{I} \\
\boldsymbol{B}= & \lambda \boldsymbol{D}_{1 x} \boldsymbol{D}_{1 x}^T \\
\boldsymbol{C}= & \boldsymbol{G}^T \boldsymbol{d}+\lambda \boldsymbol{D}_1^T\left(\boldsymbol{g}_{11}^{k-1}+\boldsymbol{g}_{21}^{k-1}+\boldsymbol{v}_1^{k-1}\right) 
 +\lambda\left(\boldsymbol{g}_{12}^{k-1}+\boldsymbol{g}_{22}^{k-1}+\boldsymbol{v}_2^{k-1}\right) \boldsymbol{D}_{1 x}^T+\alpha \boldsymbol{m}_0,
\end{aligned}\right.
\end{equation*}
where $\lambda = \mu/\gamma$ and $\alpha = \mu/\rho$. $\I$ denotes the identity matrix.

Regarding the subproblems \eqref{equation:14} and \eqref{equation:15}, their closed-form solutions can be obtained using the proximal operator of the L1-norm \cite{donoho1995noising}:
\begin{equation}
\g_{11}^k = \text{Prox}_\gamma~ ( \D_1\m^{k}-\g_{21}^{k-1}-\v_1^{k-1}) ,
\label{equation:22}
\end{equation}
\begin{equation}
\g_{12}^k = \text{Prox}_\gamma~ ( \m^{k}\D_{1x}-\g_{22}^{k-1}-\v_2^{k-1}) ,
\label{equation:23}
\end{equation}
where $\text{Prox}_\gamma(\cdot)$ is the soft-thresholding operator:
\begin{equation}
\text{Prox}_\gamma(\x)=~ \text{sign}(\x)max( |\x|-\gamma,0).
\label{equation:24}
\end{equation}
Equations \eqref{equation:16} and \eqref{equation:17} can be solved by referring to equation \eqref{equation:13}. The corresponding update formulas are:
\begin{equation}
\begin{aligned}
\g_{21}^{k}=\operatorname{Sylvester}(\boldsymbol{A}_{21}, \boldsymbol{B}_{21}, \boldsymbol{C}_{21})
\quad \text{with} \quad 
\left\{\begin{aligned}
\boldsymbol{A}_{21}= & \gamma \beta \boldsymbol{D}_1^T \boldsymbol{D}_1+\boldsymbol{I} \\
\boldsymbol{B}_{21}= & \gamma \beta\boldsymbol{D}_{1 x} \boldsymbol{D}_{1 x}^T \\
\boldsymbol{C}_{21}= &  \boldsymbol{D}_1\m^k-\boldsymbol{g}_{11}^{k}-\boldsymbol{v}_1^{k-1}
\end{aligned}\right.
\end{aligned}
\label{equation:25}
\end{equation}
\begin{equation}
\begin{aligned}
\g_{22}^{k}=\operatorname{Sylvester}(\boldsymbol{A}_{22}, \boldsymbol{B}_{22}, \boldsymbol{C}_{22})
\quad \text{with} \quad 
\left\{\begin{aligned}
\boldsymbol{A}_{22}= & \gamma\beta  \boldsymbol{D}_1^T \boldsymbol{D}_1+\boldsymbol{I} \\
\boldsymbol{B}_{22}= & \gamma \beta\boldsymbol{D}_{1 x} \boldsymbol{D}_{1 x}^T \\
\boldsymbol{C}_{22}= & \m^k\boldsymbol{D}_{1x}-\boldsymbol{g}_{12}^{k}-\boldsymbol{v}_2^{k-1}
\end{aligned}\right.
\end{aligned}
\label{equation:26}
\end{equation}
\section{Adaptive selection of regularization and parameter balancing}

\subsection{Adaptive selection of balancing parameter $\beta$ }
The balancing parameter $\beta$ controls the relative contributions of the smooth and piecewise-constant components. To adapt this balance to the current estimate, we employ an adaptive selection strategy based on robust statistics \cite{rousseeuw2018anomaly}. Robust z-scores \cite{iglewicz1993volume} identify inlier gradients and limit the influence of outlying values. For each gradient sample, the robust z-score is defined as:
\begin{equation}
\text{z-score}_i = \frac{[\g]_i-\text{median}(\g)}{\text{MAD}(\g)},
\label{equation:27}
\end{equation}
where $\g = [\D_1\m; \m\D_{1x}]$  denotes the total gradient, $[\g]_i$ denotes the \textit{i}-th sample of $\g$, and MAD denotes the median absolute deviation, defined as:
\begin{equation}
\text{MAD}(\g) = 1.4826~\text{median}(~\g-\text{median}(\g)~).
\label{equation:28}
\end{equation}
The coefficient 1.4826 ensures that the scaled $\text{MAD}$ is a consistent estimator of the standard deviation under a Gaussian distribution \cite{rousseeuw1993alternatives}.

For a threshold $\tau_{nrm}\in[2.5,4]$, if $|\text{z-score}_i|\le\tau_{nrm}$, the corresponding gradient is classified as a normal component; otherwise, it is regarded as a sparse gradient arising from a formation interface or an anomalous perturbation. The set comprising all normal gradients is denoted by:
\begin{equation}
\textit{nrm}(\g) = \{~[\g]_i:|\text{z-score}_i|\le\tau_{nrm}~\}.
\label{equation:29}
\end{equation}
Therefore, the optimal parameter $\beta^{opt}$ should satisfy the following condition: the infinity norm of the smooth-gradient component $\g_2 = [\g_{21}; \g_{22}]$ generated by the Tikhonov term is equal to that of the normal component $\textit{nrm}(\g)$ extracted from the total gradient $\g$. Accordingly, we define the function:
\begin{equation}
\phi(\beta) = \|\g_2(\beta)\|_\infty-\|nrm(\g(\beta))\|_\infty,
\label{equation:30}
\end{equation}
where $\|\cdot\|_\infty$ denotes the infinity norm.

The optimal parameter then satisfies:
\begin{equation}
\phi(\beta^{opt}) = 0.
\label{equation:31}
\end{equation}
Starting from an initial value $\beta^0$, the parameter $\beta$ is updated at iteration \textit{k} as follows:
\begin{equation}
\beta^k = \beta^{k-1}(\frac{2\|\g_2^k(\beta^{k-1})\|_\infty}{\|\g_2^k(\beta^{k-1})\|_\infty +\|nrm(\g^k(\beta^{k-1}))\|_\infty}) .
\label{equation:32}
\end{equation}
%
\subsection{Adaptive selection of regularization parameter $\lambda$}
The regularization parameter $\lambda$ enforces consistency between the model gradients and the auxiliary variables and influences the convergence behavior of the iterative process. An excessively small value may lead to substantial discrepancies between the model gradients and auxiliary variables, whereas an excessively large value may cause the model updates to be dominated by the penalty terms. To avoid reliance on a fixed parameter, we employ the RWP to select  . Because solving the model subproblem separately for each candidate value of $\lambda$ is computationally expensive, we further develop a fast computational scheme based on the discrete cosine transform (DCT) \cite{gholami2016fast}.
\subsubsection{Automatic Selection of the Parameter using RWP}
At iteration \textit{k}, we evaluate a set of logarithmically spaced candidate regularization parameters for the model subproblem in equation \eqref{equation:13}:
\begin{equation}
\lambda = [\lambda_1, \lambda_2, ...,\lambda_j].
\label{equation:33}
\end{equation}
For each candidate $\lambda_j$, the corresponding model $\m_j^k$ satisfies (21):
\begin{equation}
\m_j^k=\operatorname{Sylvester}(\boldsymbol{A}(\lambda_j), \boldsymbol{B}(\lambda_j), \boldsymbol{C}(\lambda_j)) .
\label{equation:34}
\end{equation}
The data residual for each candidate model is defined as:
\begin{equation}
\rho(\m_j^k)=\d-\G\m_j^k.
\label{equation:35}
\end{equation}
The RWP objective function can be simplified as \cite{sheriff1995exploration,lanza2013variational,aghazade2025automatic}:
\begin{equation}
\lambda^k=\underset{\lambda}{\operatorname{argmin}}~  \frac{\|FFT[(\rho(\m^k(\lambda))]\|_4^4}{\|FFT[(\rho(\m^k(\lambda))]\|_2^4},
\label{equation:36}
\end{equation}
where $FFT(\cdot)$  denotes the discrete Fourier transform.
\subsubsection{Fast Computation of Candidate Models for Parameter Selection}
Although the RWP enables automatic selection of the parameter $\lambda$, its direct implementation still requires the model subproblem to be solved separately for each candidate value. If the number of outer iterations is \textit{k} and \textit{N} candidate parameters are tested at each iteration, the direct approach requires ($\textit{k}\times \textit{N}$) solutions of the Sylvester equation, resulting in a high computational cost. To address this issue, we derive a fast expression for computing the candidate models using the discrete cosine transform.

The solution to equation \eqref{equation:13} can first be simplified as:
\begin{equation}
(\G^T\G+\alpha\I)\m_j^k+\lambda(\D_1^T\D_1\m_j^k+\m_j^k\D_{1x}\D_{1x}^T)=\R+\lambda\Q^{k-1},
\label{equation:37}
\end{equation}
where,
\begin{equation}
\left\{
\begin{aligned}
&\R= \G^T\d +\alpha\m_0,\\
&\Q^{k-1}=  \D_1^T( \g_{11}^{k-1} + \g_{21}^{k-1} +\v_1^{k-1}) + ( \g_{12}^{k-1} + \g_{22}^{k-1} +\v_2^{k-1})\D_{1x}^T.
\end{aligned}
\right.
\label{equation:38}
\end{equation}
Let $\boldsymbol{\psi}$ denote the discrete cosine transform matrix. Then,
\begin{equation}
\left\{
\begin{aligned}
&\D_1^T \D_1=\boldsymbol{\psi}^T \boldsymbol{\Sigma}_1 \boldsymbol{\psi}, \\
&\D_{1 x} \D_{1 x}^T=\boldsymbol{\psi}^T \boldsymbol{\Sigma}_{1 x} \boldsymbol{\psi},
\end{aligned}
\right.
\label{equation:39}
\end{equation}
where  $\boldsymbol{\Sigma}_1$ and  $\boldsymbol{\Sigma}_{1 x}$ are diagonal matrices including the eigenvalues. Substituting equation \eqref{equation:39} into equation \eqref{equation:37}, pre-multiplying both sides by $\boldsymbol{\psi}$, and post-multiplying both sides by $\boldsymbol{\psi}^T$ yields, after simplification:
\begin{equation}
\boldsymbol{\psi}(\G^T\G+\alpha\I)\boldsymbol{\psi}^T\tilde{\m}_j^k+\lambda(\boldsymbol{\Sigma}_1\tilde{\m}_j^k+\tilde{\m}_j^k\boldsymbol{\Sigma}_{1 x})=\tilde{\R}+\lambda\tilde{\Q}^{k-1},
\label{equation:40}
\end{equation}
where,
\begin{equation}
\left\{
\begin{aligned}
&\tilde{\m}_j^k=\boldsymbol{\psi} \m_j^k \boldsymbol{\psi}^T, \\
&\tilde{\R}=\boldsymbol{\psi} \R \boldsymbol{\psi}^T, \\
&\tilde{\Q}^{k-1}=\boldsymbol{\psi} \Q^{k-1} \boldsymbol{\psi}^T. 
\end{aligned}
\right.
\label{equation:41}
\end{equation}
We approximate the normal matrix $\G^T\G$ by a diagonal matrix in the DCT domain:
\begin{equation}
\boldsymbol{\psi} (\G^T\G +\alpha\I)\boldsymbol{\psi}^T = \Sigma_g + \alpha\I.
\label{equation:42}
\end{equation}
Because $\bm{\Sigma_1}$, $\bm{\Sigma_{1x}}$, and $\bm{\Sigma_{g}}$ are all diagonal matrices, equation \eqref{equation:42} can be solved elementwise:
\begin{equation}
\m_j^k=\text{DCT}^{-1}(\tilde{\m}_j^k)=\text{DCT}^{-1}(\frac{\tilde{\R}+\lambda_j\tilde{\Q}^{k-1}}{\bm{\Sigma_{g}}+\alpha\I+\lambda_j\bm{\Sigma_{1}}+\lambda_j\bm{\Sigma_{1x}}}),
\label{equation:43}
\end{equation}
where $\text{DCT}^{-1}(\cdot)$ denotes the inverse discrete cosine transform.

Equation \eqref{equation:43} no longer requires a separate solution of the Sylvester equation for each candidate parameter. At each iteration, approximate candidate models $\m_j^k$ are computed using DCTs and elementwise operations, and their residual whiteness scores are evaluated. After a candidate regularization parameter $\lambda_j$ is selected, the current model is updated by solving equation \eqref{equation:20}. This reduces the number of exact Sylvester-equation solves from $\textit{k}\times \textit{N}$ to \textit{k}, although the cost of candidate evaluation still increases with \textit{N}. The complete procedure for parameter selection and ADMM variable updating is summarized in Algorithm \ref{algorithm1}.

\begin{algorithm}    
\setlength{\baselineskip}{1.5em}
\caption{: Impedance inversion using the proposed ATT regularization with adaptive parameter selection}
\begin{algorithmic}
  \Statex \textbf{Input}: $\d$, $\bm{m}_0$, $k$, and hyper-parameters $\beta^0$, $\alpha$, $\xi$;
  \Statex \textbf{Initial values}: $\bm{g}_{11}^0 = \bm{g}_{12}^0 = 0$, $\bm{g}_{21}^0 = \bm{g}_{22}^0 = 0$, $\bm{v}_1^0 = \bm{v}_2^0 = 0$, $\bm{m}^0 = \bm{m}_0$.
  \While{a stopping criterion is not satisfied}
    \For{$k \leftarrow 1, 2, \cdots, \mathit{maxit}$}
      \State Estimate the $\lambda^{opt}$ from RWP using \eqref{equation:36} and \eqref{equation:43}.
      \State Update $\bm{m}^k$ using the $\lambda^{opt}$ and \eqref{equation:21}.
      \State Update $\bm{g}_{11}^k$ and $\bm{g}_{12}^k$ according to \eqref{equation:22} and \eqref{equation:23}.
      \State Update $\bm{g}_{21}^k$ and $\bm{g}_{22}^k$ according to \eqref{equation:25} and \eqref{equation:26}.
      \State Update $\bm{v}_1^k$ and $\bm{v}_2^k$ according to \eqref{equation:18} and \eqref{equation:19}.
      \State Update $\bm{\beta}^k$ according to \eqref{equation:32}.
    \EndFor
  \EndWhile
  \Statex \textbf{Output}: $\bm{z} = \exp(\bm{m}^k)$.
\end{algorithmic}
\label{algorithm1}
\end{algorithm}
\section{Examples}
We compare the Tikhonov, TV, and TT regularization methods in a single channel inversion setting where each trace is inverted independently. Then we use the ATT method where the whole traces are inverted simultaneously and thus the lateral continuity of the impedance model is taken into account. The four methods use the same forward operator, low-frequency background model, input data, and stopping criterion. The balancing parameters of TT and ATT are initialized with $\beta^0=1$, and the anisotropy parameter $\xi=2$. ATT follows Algorithm 1; the other methods use implementations based on \cite{gholami2022automatic,aghazade2024elastic,aghazade2026robust}.

For synthetic tests, the model error is evaluated using the relative model error (RME) defined below:
\begin{equation}
\text{RME}=\frac{\|\m^k-\m^*\|_F^2}{\|\m^*\|_F^2} \times 100\%,
\label{equation:44}
\end{equation}

where $\m^k$ denotes the inversion result obtained at the \textit{k}-th iteration, and $\m^*$ denotes the reference model. A smaller RME indicates closer agreement between the inversion result and the reference model.
\subsection{Overthrust model}
\begin{figure}
    \centering
    \includegraphics[width=0.28\linewidth]{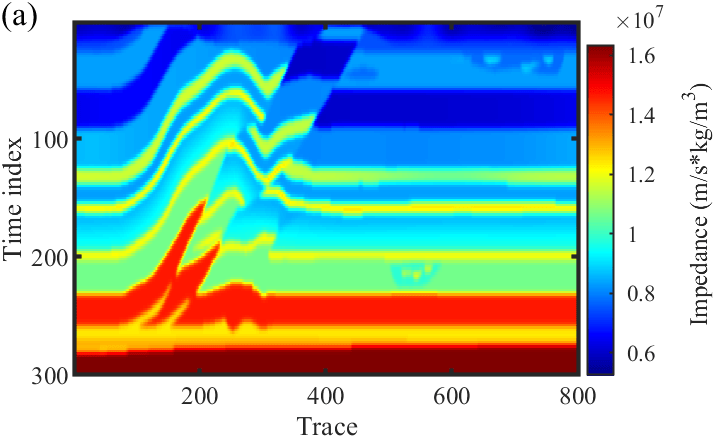}
    \includegraphics[width=0.28\linewidth]{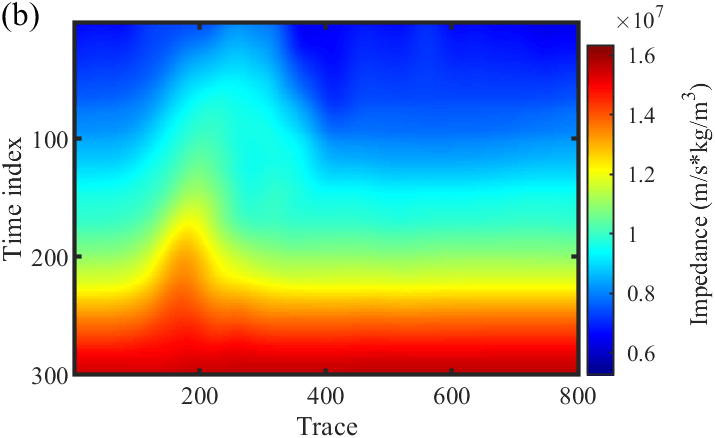}
    \includegraphics[width=0.28\linewidth]{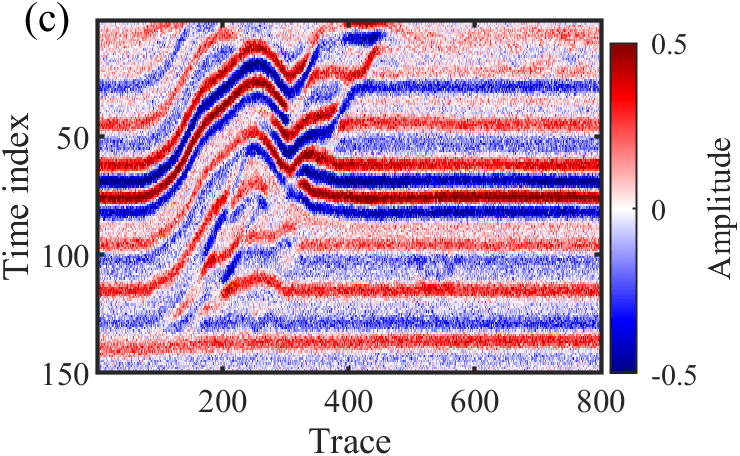}
    \caption{Overthrust model: (a) reference impedance model; (b) low frequency model; (c) noisy post-stack seismic data (SNR = 10 dB).}
    \label{fig:fig1}
\end{figure}

We first use an acoustic impedance section derived from the Overthrust model for testing. The model contains dipping strata, folds, faults, and impedance interfaces with pronounced lateral variations. It includes both abrupt interlayer impedance changes and continuous intralayer variations, making it suitable for evaluating the interface-preserving capability and lateral continuity of different methods. Fig.~\ref{fig:fig1}(a) shows the reference impedance model, and Fig.~\ref{fig:fig1}(b) shows the low-frequency model obtained by applying a low-pass filter to the reference model. Reflection coefficients were calculated from the reference impedance model and convolved with a zero-phase Ricker wavelet with a dominant frequency of 30 Hz to generate synthetic seismic records. Gaussian random noise was subsequently added to obtain the noisy seismic data with a signal-to-noise ratio (SNR) of 10 dB, as shown in Fig.~\ref{fig:fig1}(c).

\begin{figure}
    \centering
    \includegraphics[width=1\linewidth]{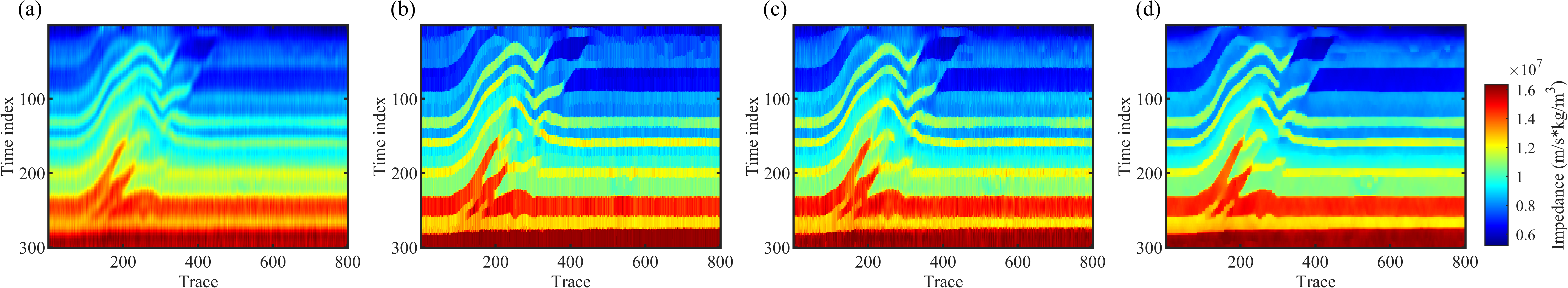}
    \includegraphics[width=1\linewidth]{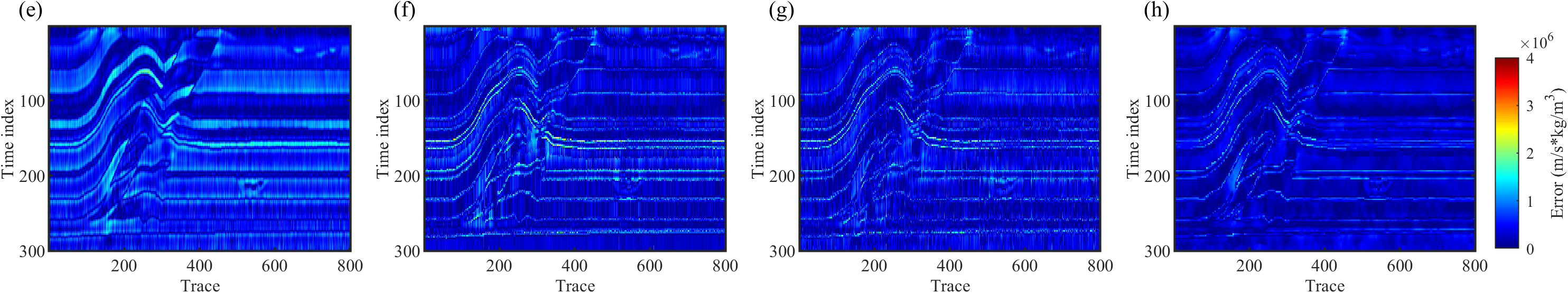}
    \caption{Inversion results and absolute error sections for the Overthrust model: (a–d) impedance results obtained using Tikhonov, TV, TT, and ATT regularization, respectively; (e–h) corresponding absolute error sections.}
    \label{fig:fig2}
\end{figure}

Fig.~\ref{fig:fig2} presents the impedance inversion results and corresponding absolute error sections obtained using the four methods. Figs.~\ref{fig:fig2}(a)–~\ref{fig:fig2}(d) show the results of Tikhonov, TV, TT, and ATT regularization, respectively, while Figs.~\ref{fig:fig2}(e)–~\ref{fig:fig2}(h) show their corresponding absolute errors. Tikhonov regularization recovers the overall impedance trend but produces blurred interfaces. TV regularization improves interface definition but introduces staircase artifacts. TT regularization provides a balance between interface preservation and continuous intralayer variations, although local lateral discontinuities remain. In comparison, ATT more accurately reconstructs the dipping strata, folds, and faults and produces the smallest error magnitudes.

\begin{figure}
    \centering
    \includegraphics[width=1\linewidth]{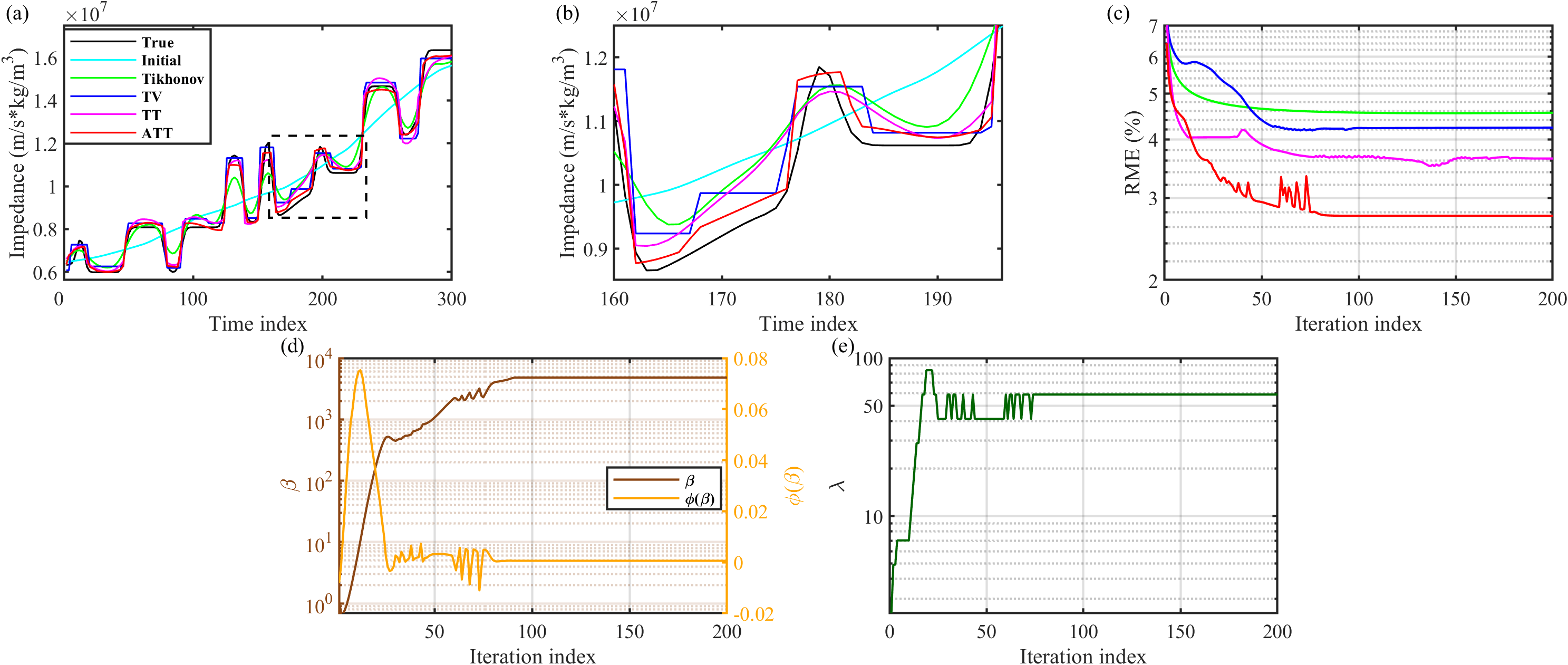}
    \caption{Inversion results and convergence behavior for the Overthrust model: (a) impedance profiles at Trace = 400; (b) enlarged view of the dashed rectangle in (a); (c) RME convergence curves for the four regularization methods; (d) convergence of the balancing parameter $\beta$ and balancing function $\phi(\beta)$ for ATT regularization; (e) adaptive updates of the regularization parameter $\lambda$.}
    \label{fig:fig3}
\end{figure}

Fig.~\ref{fig:fig3}(a) shows the single-trace impedance curves at trace 400, and Fig.~\ref{fig:fig3}(b) provides an enlarged view of the region enclosed by the dashed box. The Tikhonov result is oversmoothed at impedance discontinuities, whereas the TV result exhibits staircase artifacts. Although TT provides some improvement, local deviations from the reference curve remain. By contrast, the ATT result shows a high overall agreement with the true curve, particularly in recovering gradual impedance trends and several localized discontinuities. Fig.~\ref{fig:fig3}(c) shows the RME convergence curves of the four methods. The final RME values for Tikhonov, TV, TT, and ATT are approximately 4.6\%, 4.2\%, 3.6\%, and 2.7\%, respectively. ATT achieves the lowest final RME among the four methods. Fig.~\ref{fig:fig3}(d) shows the evolution of the balancing parameter $\beta$ and the balancing function $\phi(\beta)$ during ATT inversion. As the iterations proceed, gradually stabilizes, while converges to approximately zero after initial fluctuations. Fig.~\ref{fig:fig3}(e) presents the adaptive updates of the regularization parameter $\lambda$. Its adjustments diminish in the later iterations, when the RME curve also levels off, suggesting that parameter selection becomes consistent as the inversion approaches a stable solution.
\subsection{BP2004 model}
\begin{figure}
    \centering
    \includegraphics[width=0.3\linewidth]{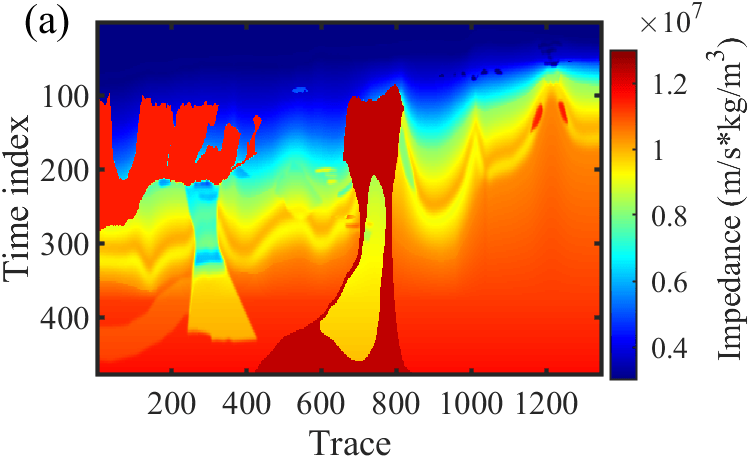}
    \includegraphics[width=0.3\linewidth]{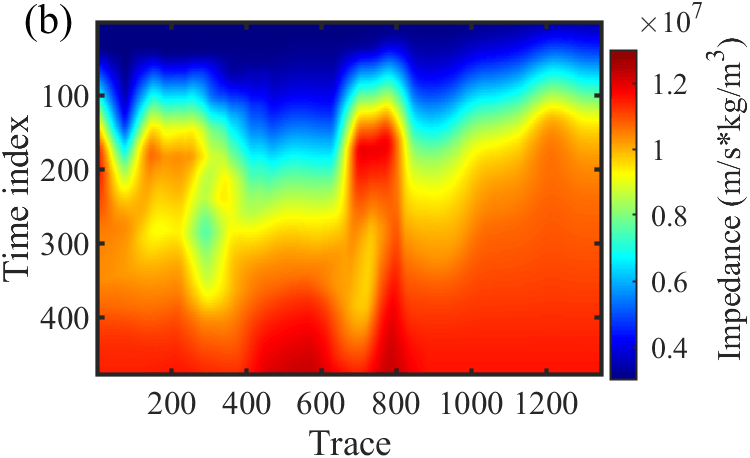}
    \includegraphics[width=0.3\linewidth]{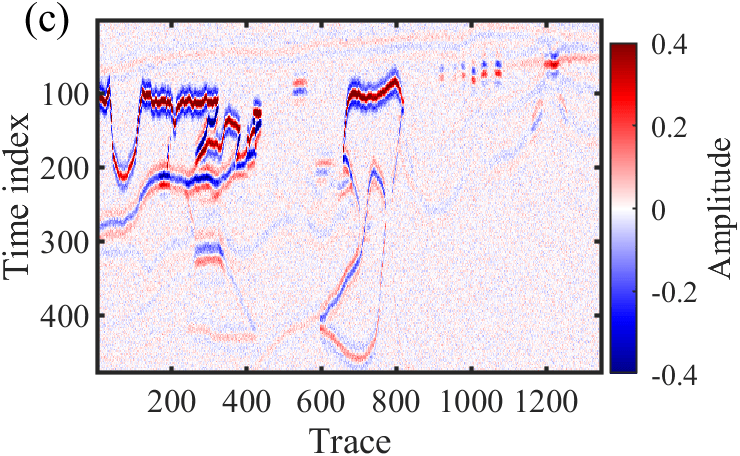}
    \caption{BP2004 model and input data: (a) reference impedance model; (b) low-frequency model; (c) noisy post-stack seismic data (SNR = 10 dB).}
    \label{fig:fig4}
\end{figure}

To further evaluate the ability of the different methods to handle strong impedance contrasts and complex boundaries, the BP2004 model \cite{billette20052004} was used for testing. Compared with the Overthrust model, this model contains multiple high impedance anomalies, steeply dipping boundaries, and rapid lateral variations, thereby imposing greater demands on impedance interface and amplitude recovery. The wavelet, noise level, and background-model construction procedure are the same as in Section IV.A. Figs.~\ref{fig:fig4}(a)–~\ref{fig:fig4}(c) show the reference impedance model, the low-frequency model, and the noisy seismic data with an SNR of 10 dB, respectively.

\begin{figure}
    \centering
    \includegraphics[width=1\linewidth]{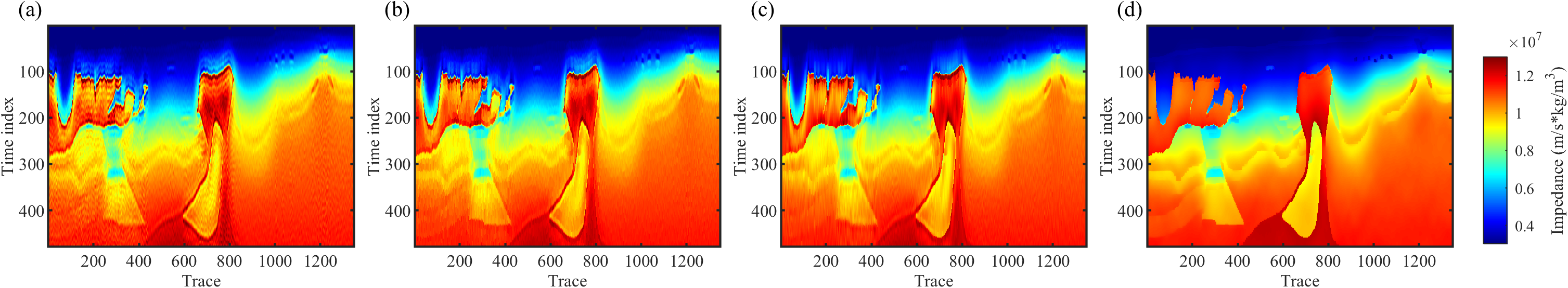}
    \includegraphics[width=1\linewidth]{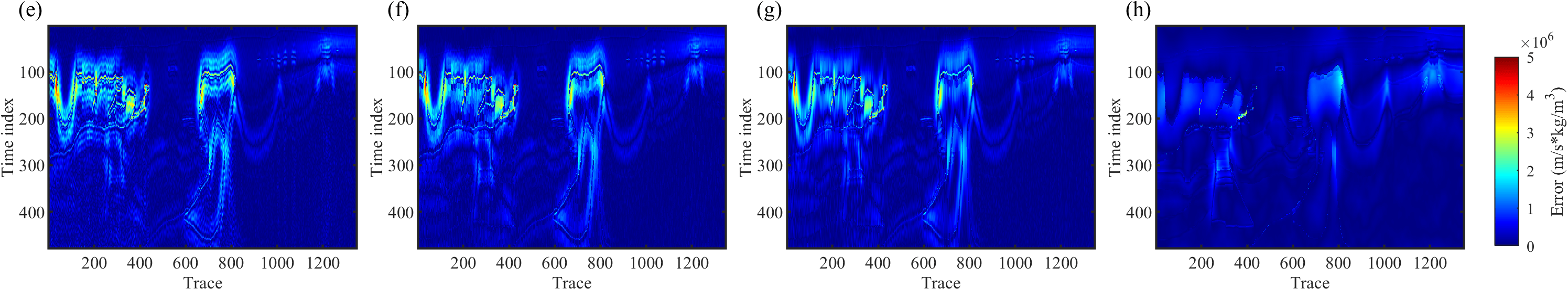}
    \caption{Inversion results and absolute-error sections for the BP2004 model: (a–d) impedance results obtained using Tikhonov, TV, TT, and ATT regularization, respectively; (e–h) corresponding absolute-error sections.}
    \label{fig:fig5}
\end{figure}

Fig.~\ref{fig:fig5} presents the inversion results and corresponding absolute errors obtained using the four methods. All methods recover the principal high impedance anomalies and their overall structure, but clear differences are observed near steeply dipping boundaries and strong impedance discontinuities. The Tikhonov and TV results contain relatively strong local oscillations or boundary errors. TT suppresses these errors to some extent, but the lateral continuity of some steeply dipping interfaces remains unstable. The ATT result has a relatively continuous overall structure, and its error range is also noticeably reduced.

\begin{figure}
    \centering
    \includegraphics[width=1\linewidth]{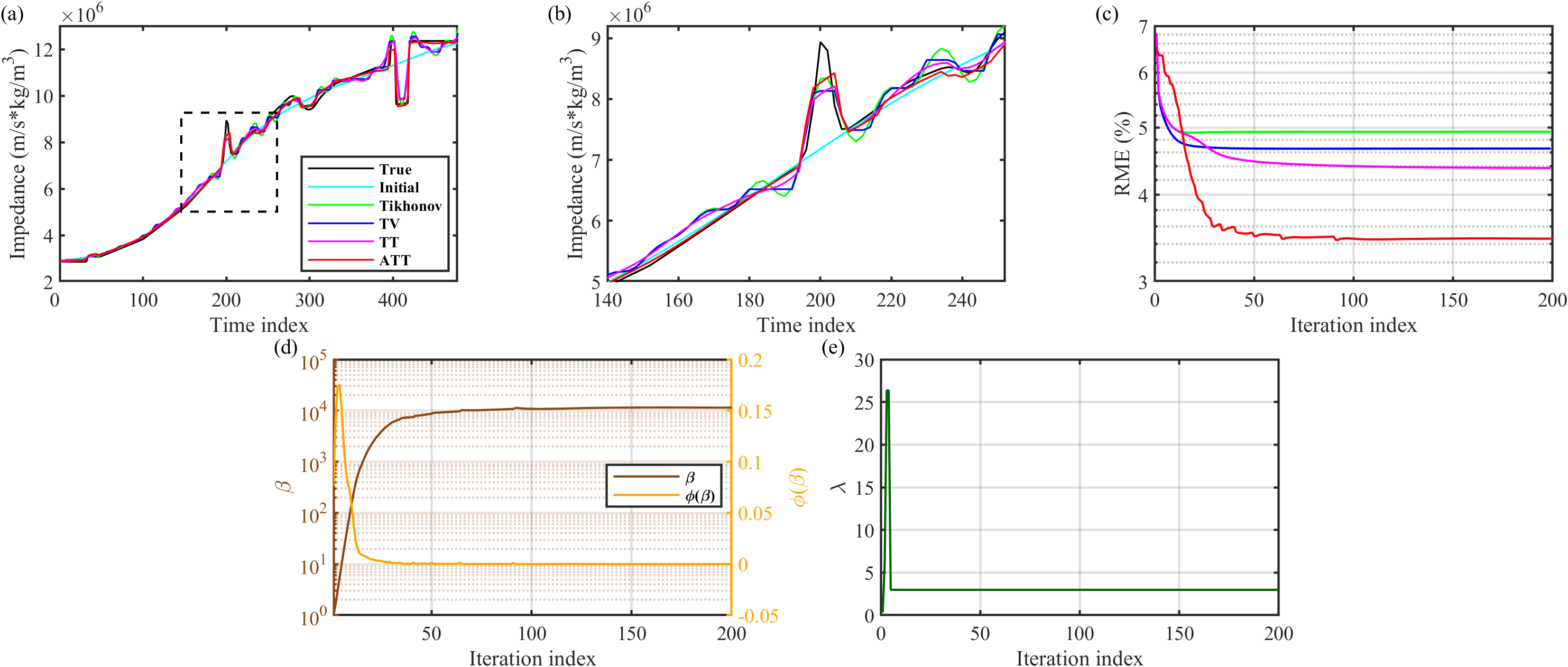}
    \caption{Inversion results and convergence behavior for the BP2004 model: (a) impedance profiles at Trace = 600; (b) enlarged view of the dashed rectangle in (a); (c) RME convergence curves for the four regularization methods; (d) convergence of the balancing parameter $\beta$ and balancing function $\phi(\beta)$ for ATT regularization; (e) adaptive updates of the regularization parameter $\lambda$.}
    \label{fig:fig6}
\end{figure}

Fig.~\ref{fig:fig6}(a) shows the impedance curves at trace 600, and Fig.~\ref{fig:fig6}(b) provides an enlarged view of the selected region. All four methods recover the main impedance trend, but underestimate the amplitude to varying degrees near the strong discontinuity at approximately time sample = 200. ATT reduces the oscillations on both sides of the abrupt interface and improves the inversion performance in the adjacent gradually varying intervals. Fig.~\ref{fig:fig6}(c) shows that the RME values of Tikhonov, TV, TT, and ATT are approximately 4.9\%, 4.7\%, 4.4\%, and 3.5\%, respectively. The larger errors relative to the Overthrust example are consistent with the difficulty of recovering the stronger contrasts and more complex boundaries. In Fig.~\ref{fig:fig6}(d), the balancing parameter  $\beta$ gradually stabilizes after an initial adjustment, while $\phi(\beta)$ rapidly approaches zero. Similarly, Fig.~\ref{fig:fig6}(e) shows that the regularization parameter $\lambda$, selected using the residual whiteness principle, changes during the early iterations before gradually stabilizing. Local adjustments in both parameters broadly coincide with the stagewise decreases in the ATT error curve, indicating that the adaptive weighting strategy remains effective for complex models.
\subsection{Field data application}
\begin{figure}
    \centering
    \includegraphics[width=0.35\linewidth]{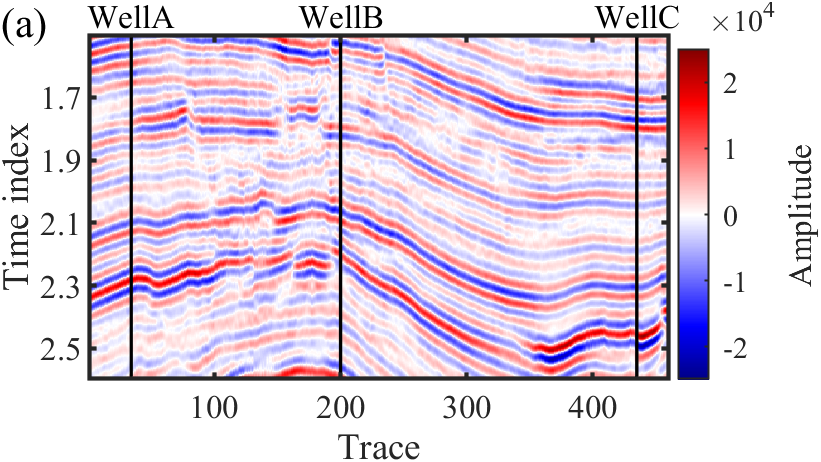}
    \includegraphics[width=0.35\linewidth]{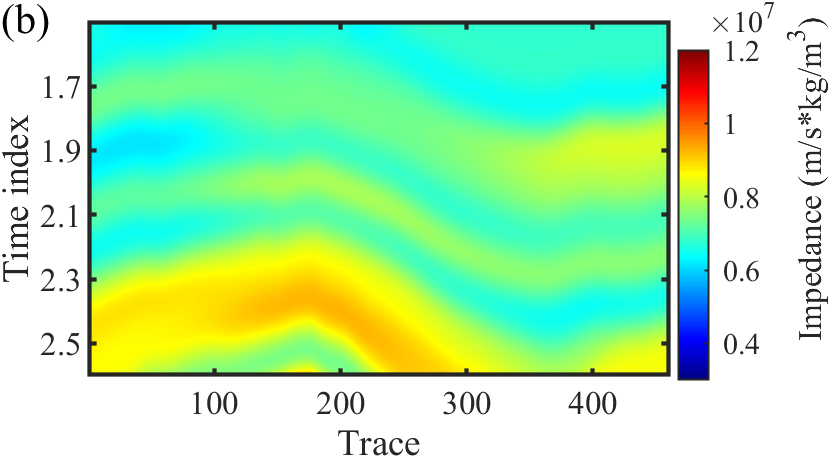}
    \caption{Field-data: (a) observed post-stack seismic section; (b) low-frequency model.}
    \label{fig:fig7}
\end{figure}

Here, the proposed method is further applied to field data from a study area to evaluate its practical effectiveness. As shown in Fig.~\ref{fig:fig7}(a), the selected post-stack seismic section consists of 460 traces, each containing 550 samples with a sampling interval of 2 ms. The black vertical lines indicate the locations of three vertical wells. Fig.~\ref{fig:fig7}(b) shows the low-frequency background model used for the inversion. The Tikhonov, TV, TT, and proposed ATT methods were applied separately, and their inversion results were evaluated against the impedance logs from the three wells.

\begin{figure}
    \centering
    \includegraphics[width=1\linewidth]{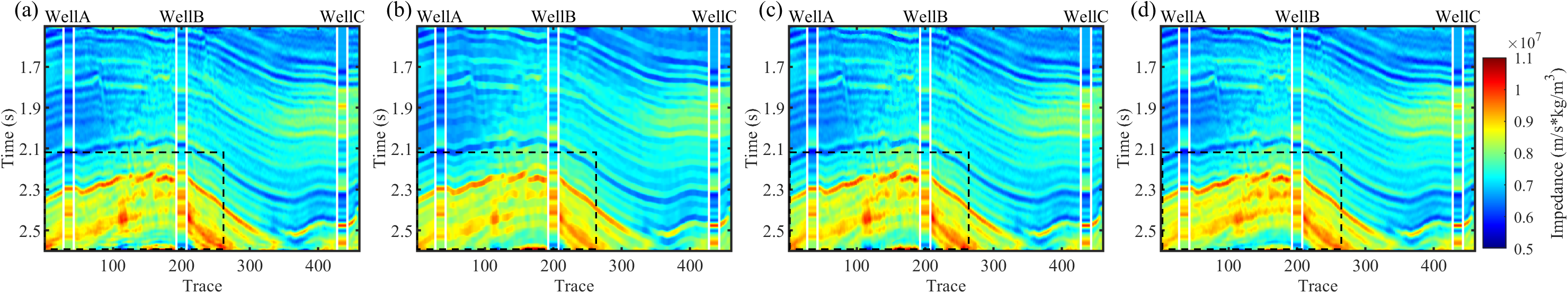}
    \caption{Field-data impedance inversion results obtained using (a) Tikhonov, (b) TV, (c) TT, and (d) ATT regularization. Well impedance logs are superimposed as narrow color strips at their corresponding locations.}
    \label{fig:fig8}
\end{figure}

\begin{figure}
    \centering
    \includegraphics[width=1\linewidth]{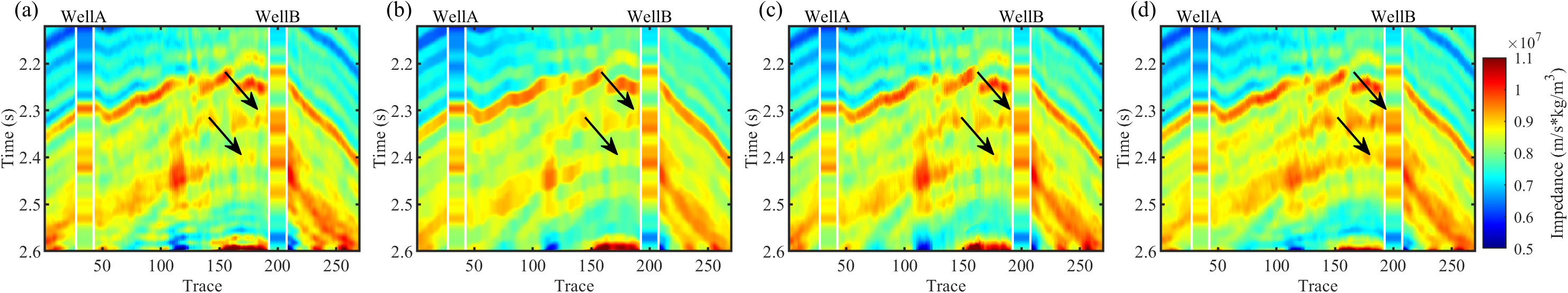}
    \caption{Enlarged views of the regions enclosed by the dashed boxes in Fig. 8: (a) Tikhonov; (b) TV; (c) TT; (d) ATT. Arrows indicate locations used for comparing local impedance structures.}
    \label{fig:fig9}
\end{figure}

Figs.~\ref{fig:fig8}(a)–\ref{fig:fig8}(d) show the impedance inversion results obtained using the four methods, respectively, with the well logs superimposed at their corresponding locations in narrow color strips. Overall, all four methods recover the principal impedance distribution of the section, although differences remain in interface definition and lateral continuity. For a clearer comparison, the regions enclosed by the black dashed boxes in Fig.~\ref{fig:fig8} are enlarged and presented in Fig.~\ref{fig:fig9}. Near the locations indicated by the arrows, ATT provides a clearer delineation of the local interfaces and the impedance variations on both sides, showing better agreement with the trends observed in the nearby well logs.
\begin{figure}
    \centering
    \includegraphics[width=1\linewidth]{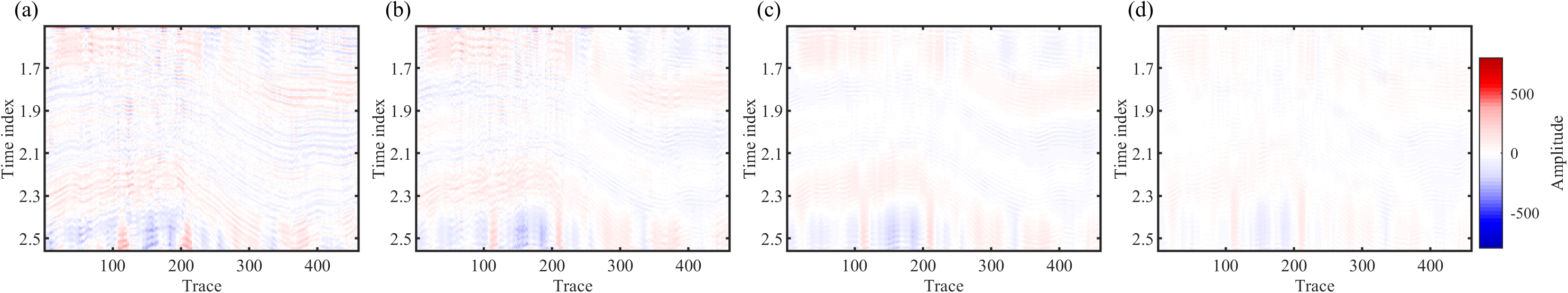}
    \caption{Differences between the synthetic and observed post-stack seismic records from the impedance obtained using (a) Tikhonov, (b) TV, (c) TT, and (d) ATT.}
    \label{fig:fig10}
\end{figure}
\begin{figure}
    \centering
    \includegraphics[width=0.7\linewidth]{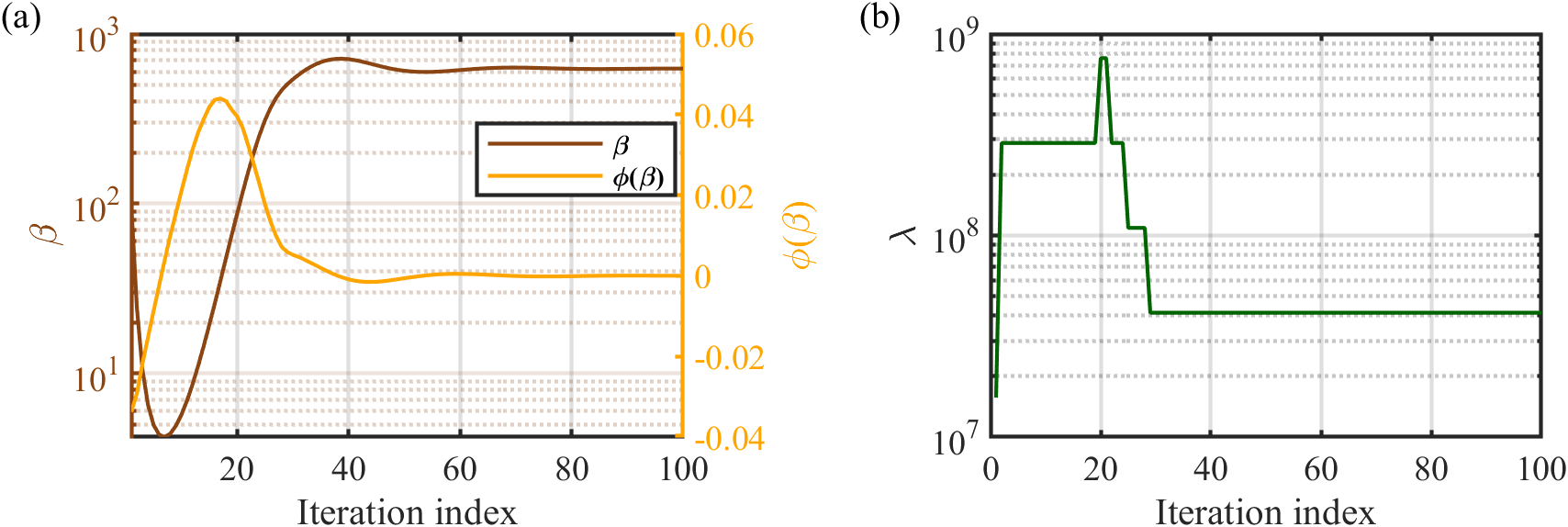}
    \caption{Parameter evolution in ATT inversion of the field data: (a) the balancing parameter  $\beta$ and balancing function $\phi(\beta)$; (b) the regularization parameter $\lambda$.}
    \label{fig:fig11}
\end{figure}

In addition, synthetic seismic records were generated from the inverted impedance models shown in Fig.~\ref{fig:fig8} and compared with the original post-stack seismic record in Fig.~\ref{fig:fig7}(a). The corresponding difference sections are shown in Fig.~\ref{fig:fig10}. Layered residuals remain visible in the difference sections obtained from the Tikhonov, TV, and TT results, whereas the ATT result exhibits generally weaker residuals, indicating a better fit between the inverted model and the observed seismic data. Fig.~\ref{fig:fig11} shows the corresponding parameter updates: $\beta$ stabilizes after initial adjustments as $\phi(\beta)$ approaches zero, while $\lambda$ remains stable after several stepwise changes during the first 30 iterations. These results indicate that the automatic parameter selection procedures reach stable values in the field-data inversion.
\section{Discussion}
\subsection{Robustness of the selection of balancing parameter $\beta$}
\begin{figure}
    \centering
    \includegraphics[width=1\linewidth]{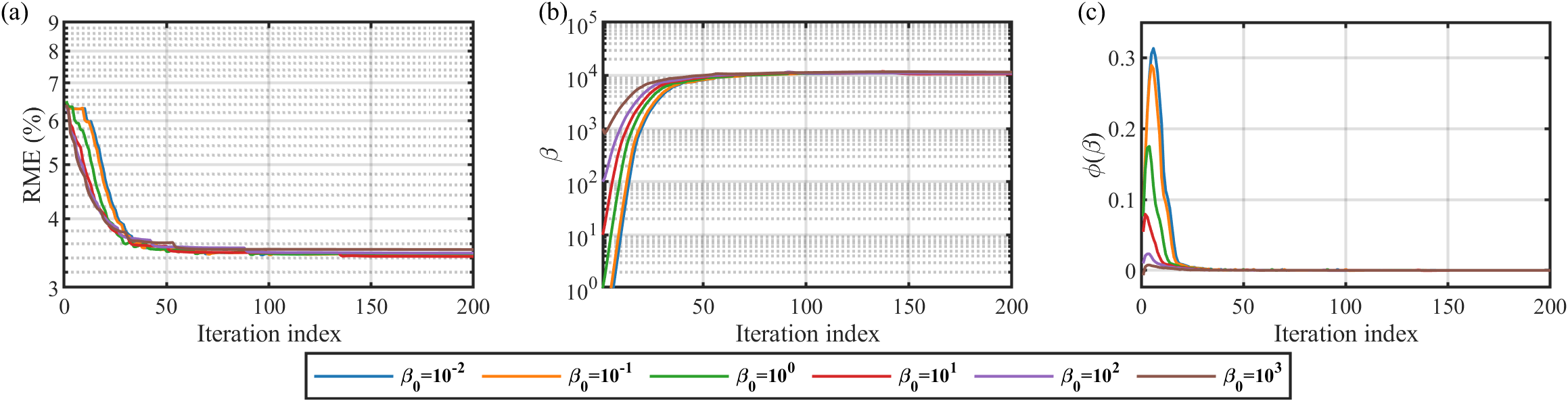}
    \caption{Convergence behavior of the adaptive balancing parameter for different initial values: (a) RME; (b) balancing parameter $\beta$; (c) balancing function $\phi(\beta)$.}
    \label{fig:fig12}
\end{figure}

The balancing parameter $\beta$ controls the relative contributions of the smooth and piecewise-constant components in ATT regularization, and its value directly affects the balance between background impedance recovery and boundary preservation. To assess the sensitivity of the adaptive strategy to the initial value $\beta^0$, we used the BP2004 model described in Section IV.B and set  $\beta^0 = \{10^{-2}, 10^{-1}, 10^{0}, 10^{1}, 10^{2}, 10^{3}\}$, while keeping the seismic data, low-frequency model, noise level, and other inversion parameters unchanged. Fig.~\ref{fig:fig12} shows the iterative evolution of the RME, $\beta$, and $\phi(\beta)$ for different initial values of $\beta^0$. In Fig.~\ref{fig:fig12}(a), the RME curves exhibit similar overall trends. They initially decrease to approximately 6.3\%, decline further between iterations 20 and 50, and eventually stabilize at approximately 3.4\%. Smaller initial values slightly delay the early reduction in error but do not appreciably affect the final error. Fig.~\ref{fig:fig12}(b) shows that $\beta$ gradually increases from the different initial values and approaches $10^{4}$ after a brief adjustment. Correspondingly, $\phi(\beta)$ in Fig.~\ref{fig:fig12}(c) exhibits positive peaks of different magnitudes during the early iterations, rapidly approaches zero and remains close to zero during the later iterations. The stagewise reduction in RME broadly coincides with the adjustments in $\beta$ and $\phi(\beta)$, indicating that the adaptive update progressively corrects the imbalance between the two gradient components. Overall, the initial value mainly affects the early convergence rate and parameter-evolution path, but has little influence on the accuracy of the final inversion result.

\begin{figure}
    \centering
    \includegraphics[width=1\linewidth]{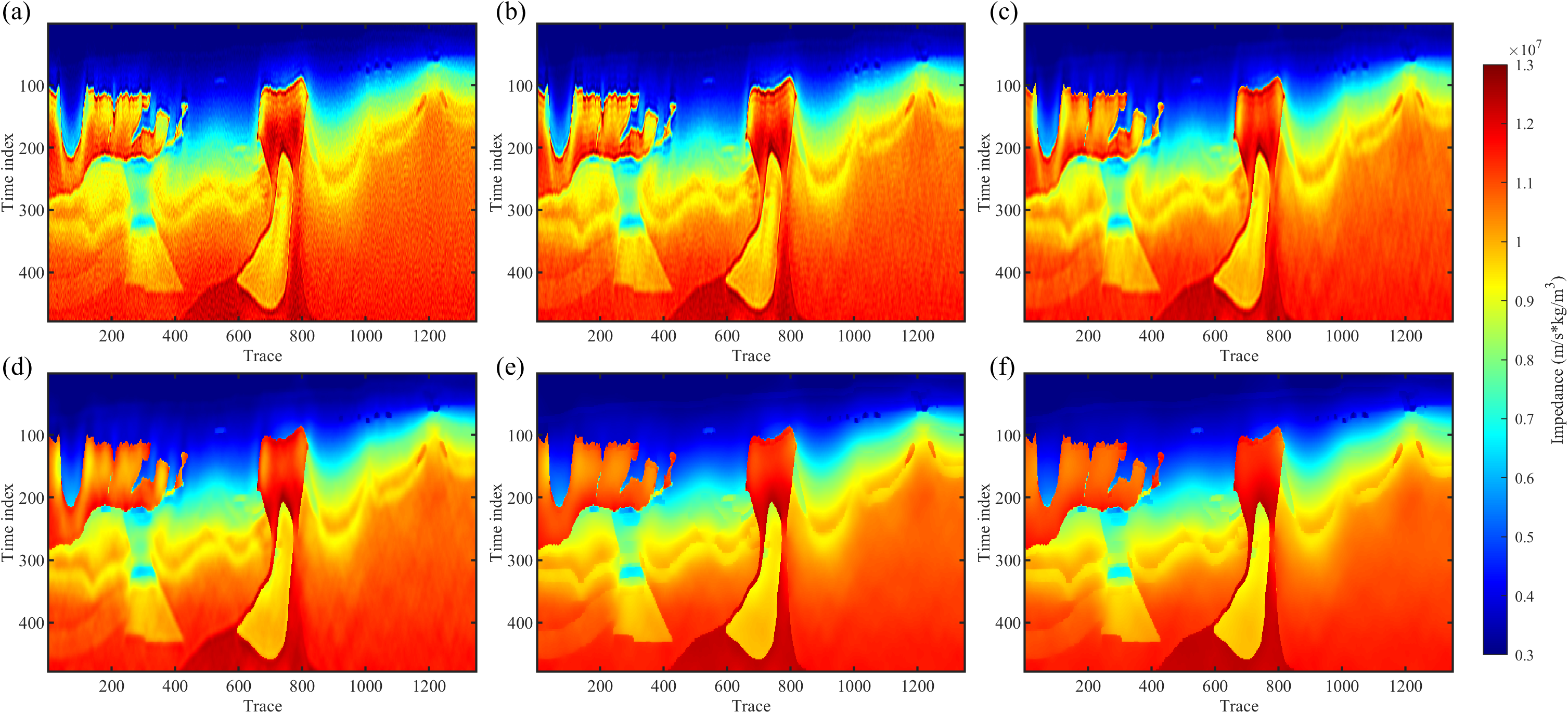}
    \caption{Inversion results obtained with fixed and adaptive balancing parameters: (a)$\beta=10^0$; (b)$\beta=10^1$; (c)$\beta=10^2$; (d)$\beta=10^3$; (e)$\beta=10^4$; (f) adaptive $\beta$.}
    \label{fig:fig13}
\end{figure}

\begin{figure}
    \centering
    \includegraphics[width=1\linewidth]{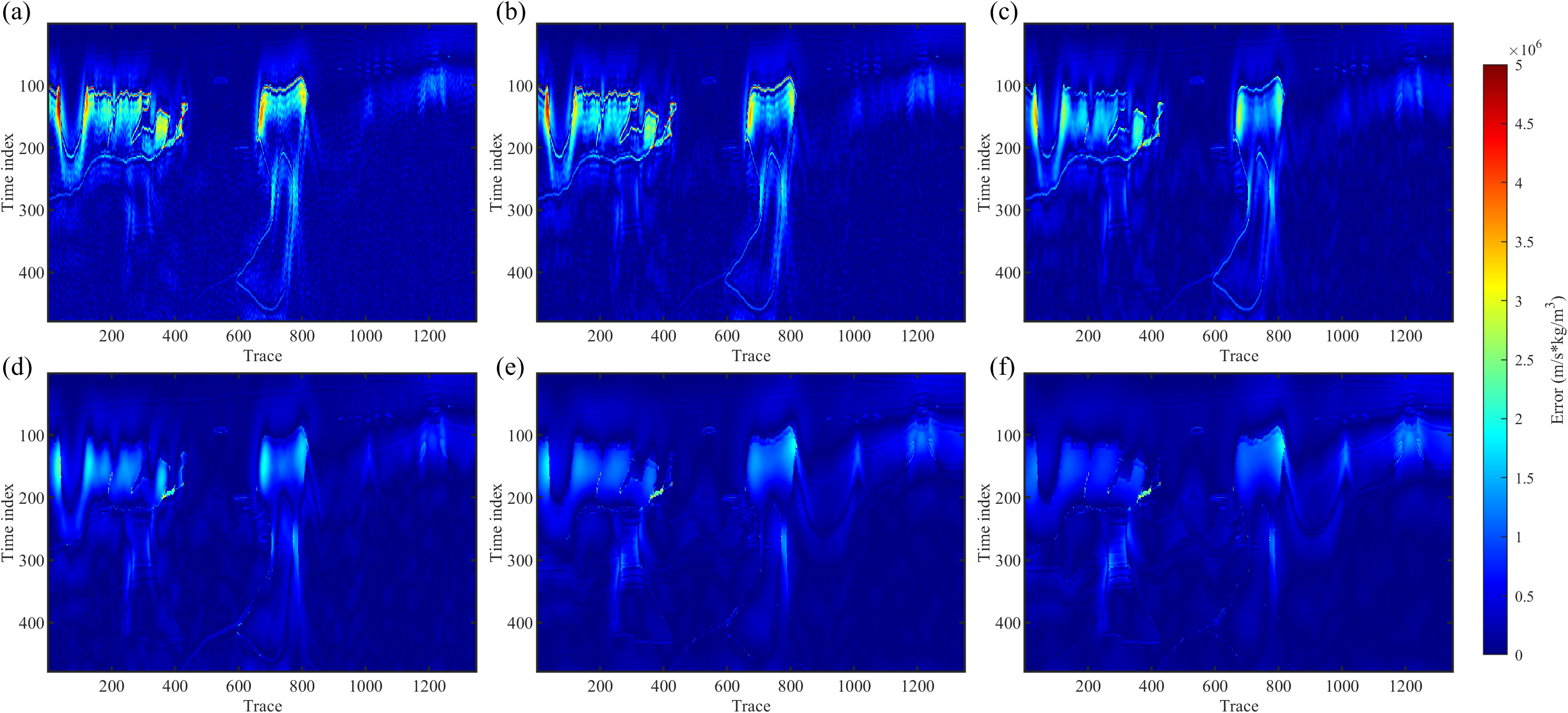}
    \caption{Absolute errors corresponding to the inversion results in Fig. 13: (a)$\beta=10^0$; (b)$\beta=10^1$; (c)$\beta=10^2$; (d)$\beta=10^3$; (e)$\beta=10^4$; (f) adaptive $\beta$.}
    \label{fig:fig14}
\end{figure}

To further evaluate the differences between fixed and adaptive values of $\beta$, Figs.~\ref{fig:fig13} and ~\ref{fig:fig14} show the inversion results and corresponding absolute errors, respectively, for fixed $\beta=\{10^0, 10^1, 10^2, 10^3, 10^4$\}, as well as for the adaptive $\beta$. When $\beta$ =$10^{0}$ or $10^{1}$, the results contain vertical oscillations and localized noise, with errors concentrated mainly around the upper high impedance anomalies and steeply dipping boundaries. Increasing $\beta$ to $10^{2}$ suppresses some of these oscillations, although spatially continuous boundary errors remain near strong impedance discontinuities. At $\beta$ =$10^{3}$, the background becomes more stable and the overall error is substantially reduced. Further increasing $\beta$ to $10^{4}$ gives the smallest error among the tested fixed values. Nevertheless, the adaptive $\beta$ result further reduces residual background artifacts and localized boundary errors while maintaining the main impedance structures.

\begin{figure}
    \centering
    \includegraphics[width=0.7\linewidth]{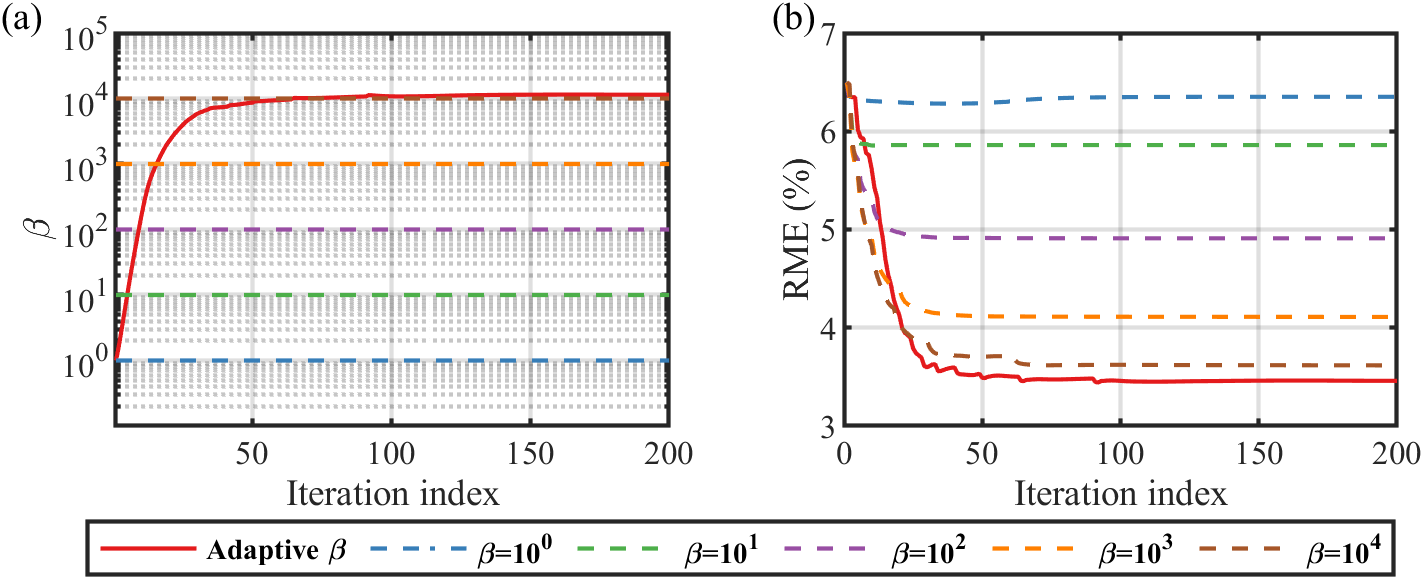}
    \caption{Comparison of fixed and adaptive balancing parameters: (a) parameter trajectories; (b) RME convergence curves.}
    \label{fig:fig15}
\end{figure}

Fig.~\ref{fig:fig15} further quantifies the effects of the different parameter selection strategies. In Fig.~\ref{fig:fig15}(a), the adaptive $\beta$ gradually increases from its initial value of $10^{0}$ and stabilizes at approximately $10^{4}$ after the 60th iteration. In Fig.~\ref{fig:fig15}(b), the RME decreases from approximately 6.3\% to 5.9\%, 4.9\%, 4.1\%, and 3.6\% as the fixed value increases from $10^{0}$ to $10^{4}$. By comparison, the adaptive method achieves an RME of approximately 3.4\%, which is lower than those obtained using all tested fixed values. Thus, although its final parameter value is close to the largest tested fixed value, adaptive updating yields a lower model error. Together with Fig.~\ref{fig:fig12}, these results demonstrate that the adaptive update is stable over the tested range of initial values and reduces the dependence on fixed parameter scanning. Nevertheless, this conclusion is limited to the present model and parameter range. Extremely large or small initial values should still be avoided in practical applications, and the other regularization parameters should be selected appropriately.
\subsection{Performance and computational efficiency of automatic regularization parameter selection}
\begin{figure}
    \centering
    \includegraphics[width=1\linewidth]{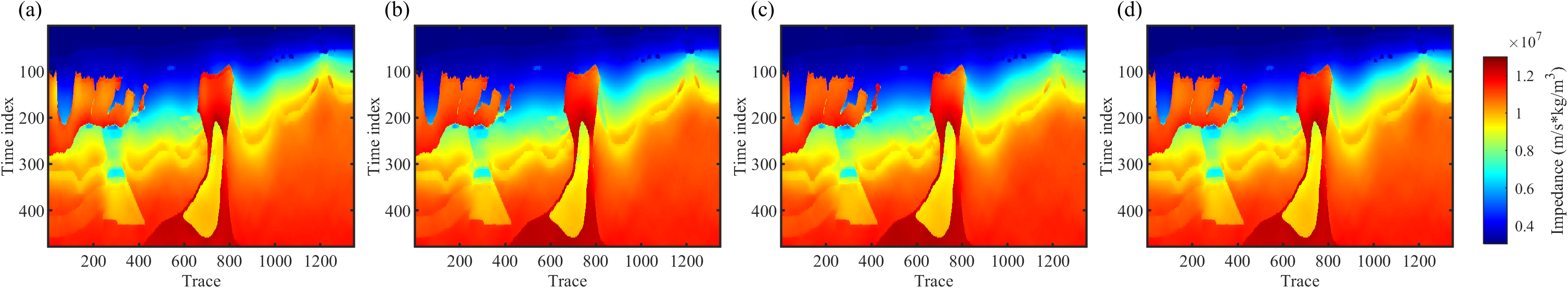}
    \caption{Inversion results obtained with fixed and adaptive regularization parameters: (a)$\lambda=10^0$; (b)$\lambda=10^1$; (c)$\lambda=2\times10^1$; (d)adaptive $\lambda$.}
    \label{fig:fig16}
\end{figure}

\begin{figure}
    \centering
    \includegraphics[width=1\linewidth]{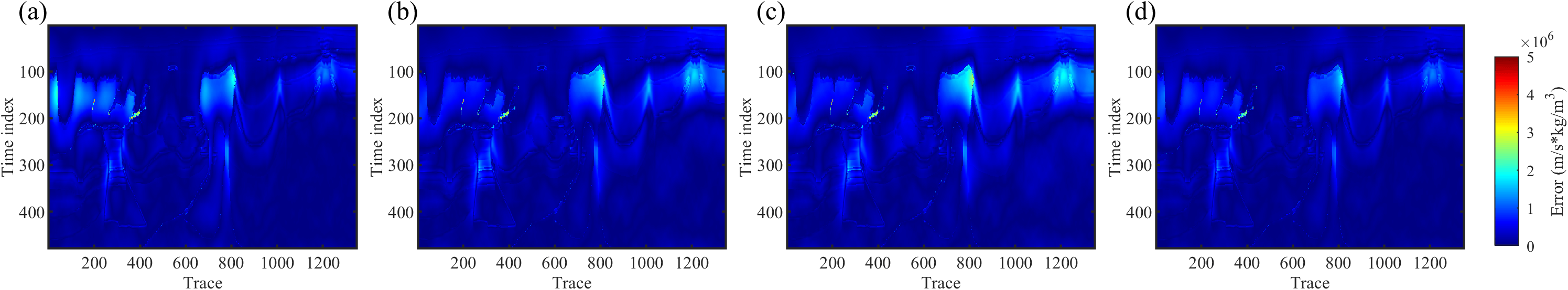}
    \caption{Absolute errors corresponding to the inversion results in Fig. 16: (a)$\lambda=10^0$; (b)$\lambda=10^1$; (c)$\lambda=2\times10^1$; (d)adaptive $\lambda$.}
    \label{fig:fig17}
\end{figure}

To evaluate the influence of the regularization parameter   on the inversion, tests are conducted using fixed values of $10^{0}$, $10^{1}$, and $2\times 10^{1}$, together with the adaptive selection strategy. Fig.~\ref{fig:fig16} presents the corresponding impedance inversion results, and Fig.~\ref{fig:fig17} shows their absolute errors. All four cases recover the main impedance structures of the model. However, the local errors gradually increase as the fixed value of $\lambda$ increases. In particular, when $\lambda=2\times 10^{1}$, relatively significant errors occur along the boundaries of the high-impedance bodies and in their surrounding regions. The result obtained with $\lambda=10^{0}$ is closer to the adaptive result, although the latter still exhibits slightly smaller local errors.

\begin{figure}
    \centering
    \includegraphics[width=0.7\linewidth]{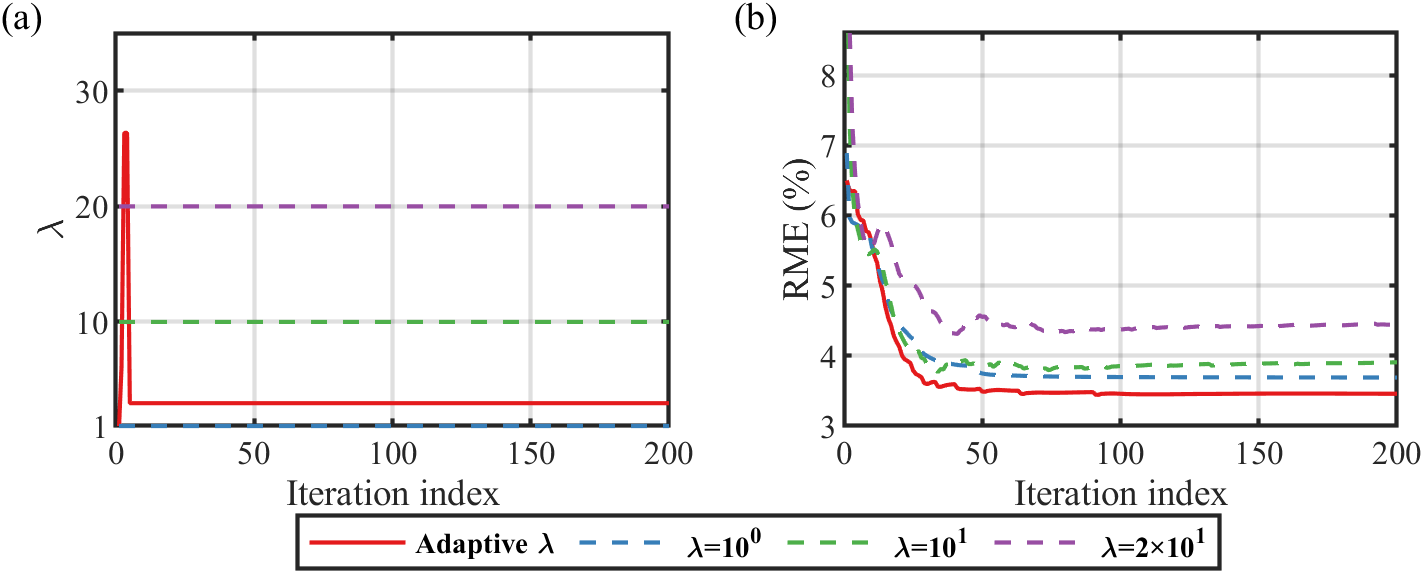}
    \caption{Comparison of fixed and adaptive regularization parameters: (a) parameter trajectories; (b) RME convergence curves.}
    \label{fig:fig18}
\end{figure}

Fig.~\ref{fig:fig18} compares the fixed and adaptive regularization parameters $\lambda$ and their corresponding RME convergence behavior. In Fig.~\ref{fig:fig18}(a), the dashed lines represent the three fixed $\lambda$ values, whereas the solid line shows the evolution of the adaptive $\lambda$. After an initial spike, the adaptive regularization parameter value decreases rapidly and stabilizes near 3. Fig.~\ref{fig:fig18}(b) shows that the fixed value $\lambda=2\times 10^{1}$ produces the highest RME, followed by $\lambda=10^{1}$. Although the fixed value $\lambda=10^{0}$ already yields a relatively low error, the adaptive strategy further reduces the RME slightly. These results indicate that the automatic selection strategy can achieve favorable inversion results within a finite number of iterations without requiring an appropriate fixed value to be specified in advance.

\begin{figure}
    \centering
    \includegraphics[width=1\linewidth]{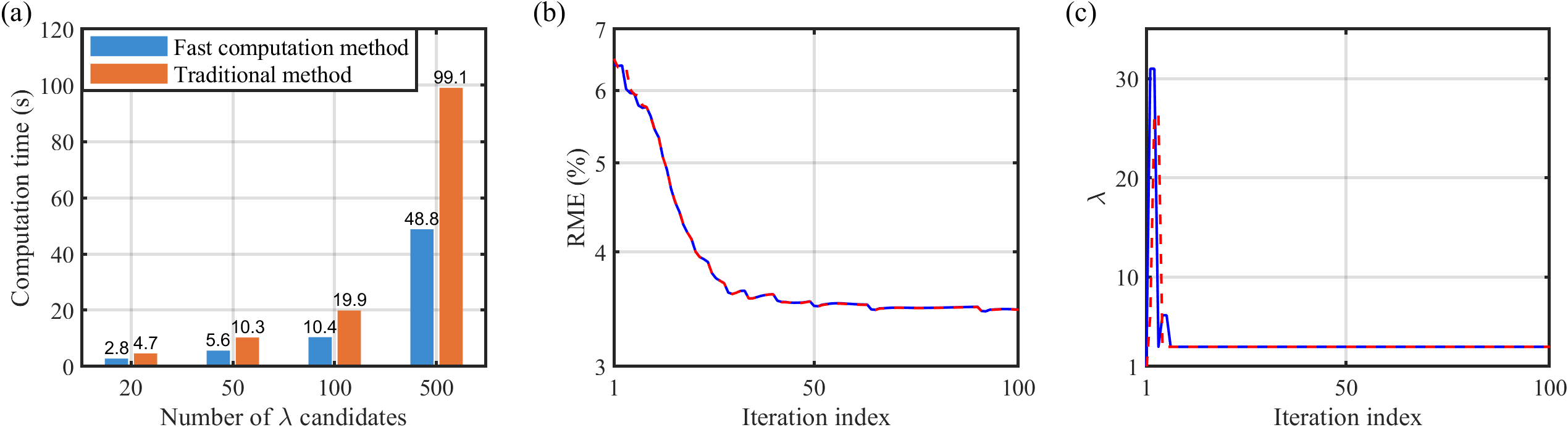}
    \caption{Comparison of the fast candidate-evaluation and the conventional method: (a) computation time per outer iteration for different numbers of candidate parameters; (b) RME convergence curves; (c) adaptive updates of the regularization parameter $\lambda$.}
    \label{fig:fig19}
\end{figure}

To evaluate the computational efficiency of the fast method and its influence on the inversion process, it was compared with the conventional method. Here, the conventional method refers to separately solving the Sylvester equation for each candidate $\lambda$ at every outer iteration, computing the corresponding data residuals, and selecting the $\lambda$ that minimizes the RWP criterion. Both methods use the same candidate set, initial model, initial balancing parameter $\beta^0$, and iteration settings. Fig.~\ref{fig:fig19}(a) shows the computation time per outer iteration. As the number of candidate parameters increases from 20 to 500, the computation time of the conventional method increases from 4.7 to 99.1 s, whereas that of the fast method increases from 2.8 to 48.8 s. Although the computation time of both methods increases with the number of candidates, the fast method exhibits a slower rate of increase and reduces the computation time by approximately 50.8\% for 500 candidates, demonstrating the efficiency gained by avoiding repeated solutions of the Sylvester equation.

Figs.~\ref{fig:fig19}(b) and~\ref{fig:fig19}(c) show the RME convergence curves and the trajectories of $\lambda$, respectively. The two RME curves nearly overlap and both eventually stabilize at approximately 3.5\%. Although the $\lambda$ trajectories differ during the initial iterations, they subsequently converge and stabilize at similar values. These results indicate that the fast computation does not appreciably alter the overall convergence behavior or final inversion accuracy. Overall, the fast method reduces the computational cost of parameter searching while maintaining inversion performance comparable to that of the conventional method, with its efficiency advantage becoming more pronounced as the number of candidate parameters increases.
\section{Conclusions}
We developed a post-stack acoustic impedance inversion method using anisotropic Tikhonov–TV regularization and automatic parameter selection. Considering that subsurface impedance exhibits both interface discontinuities and gradual intralayer variations in the vertical and horizontal directions, the proposed method imposes TV and Tikhonov constraints on the piecewise-constant and smooth components, respectively, to improve the reconstruction of local boundaries and inter-trace structures. Because the decomposition of these two components and the convergence behavior of ADMM are governed by the balancing and regularization parameters, respectively, the balancing parameter is adaptively updated according to robust statistical characteristics of the model gradients, whereas the regularization parameter is selected based on residual whiteness. A fast candidate-solution computation scheme is further introduced to reduce the computational cost of parameter selection. Parameter comparison experiments show that inappropriate fixed values degrade the inversion results, whereas the automatic strategy reduces their sensitivity to the initial balancing parameter. Finally, synthetic model and field data tests demonstrate that the proposed method recovers complex subsurface impedance structures more effectively than the Tikhonov, TV, and TT methods.

\bibliography{references}  

\end{document}